\documentclass[twocolumn,secnumarabic,amssymb,floatfix,aps]{revtex4-2}

\usepackage{graphicx}
\usepackage{dcolumn}
\usepackage{bm,amsmath,amsbsy}     
\usepackage{braket}
\usepackage{appendix}
\usepackage{xcolor}
\usepackage{array}
\usepackage{booktabs}
\usepackage{soul}
\setstcolor{red}
\usepackage{amsthm}
\usepackage{amsfonts}
\usepackage{float}
\usepackage{pstricks}           
\usepackage{mathrsfs}
\usepackage{hyperref}           
\usepackage{hypernat}           
\usepackage[latin1]{inputenc}   
\usepackage{subfigure}
\usepackage{slashed}    
\usepackage{color}
\usepackage{soul}

\begin{document}

\title{Highly accurate semiclassical strong-field Herman-Kluk propagator method for high-harmonic generation}

\author{Phi-Hung Tran}
\author{Hao Quan Truong}
\author{R. Esteban Goetz}
\author{Anh-Thu Le}
\affiliation{Department of Physics, University of Connecticut, 196A Auditorium Road, Unit 3046, Storrs, CT 06269}

\date{\today}

\begin{abstract}
We extend our recently developed semiclassical strong-field Herman-Kluk propagator (SFHK) method to calculate high-order harmonic generation (HHG) for atoms in intense lasers. We show that our method, based on a combination of the Herman-Kluk propagator and the strong-field approximation, can provide highly accurate results for both HHG yield and phase, nearly identical to those from the exact numerical solutions of the time-dependent Schr\"odinger equation. We provide detailed analyses of our method and its applications to the HHG process, particularly the recombination time. The main computational task in this method is to solve the classical Newton equations for the active electron in the combined atomic potential and laser-electron interaction. The motion of the centers of the electron wave packets, modeled by coherent states, is governed by independent classical trajectories so that the computation can therefore be parallelized very efficiently.    
\end{abstract}

\maketitle

\section{Introduction}

High-order harmonic generation (HHG) from atoms and molecules in intense lasers has been studied extensively both experimentally and theoretically since its discovery in late 1980s. At the conceptual level, a HHG process can be understood using the three-step model \cite{Krause:prl92, Corkum:prl93}: first, the active electron in the target is released into the continuum by tunneling; second, it is accelerated by the laser electric field and might be driven back to the parent ion; and third, the electron recombines with the parent ion to emit a high-energy photon. A semiclassical formulation of the three-step model based on the strong-field approximation (SFA) is given by Lewenstein {\it et al} \cite{Lewenstein:pra94}. A major problem with the Lewenstein model is that the interaction between electron in the continuum with the target ion is completely neglected. Therefore, in the past two decades, there have been various attempts to improve the model by including this interaction in different ways. However, despite these efforts, the agreements with the exact numerical solutions of the time-dependent Schr\"odinger equation (TDSE) are still mostly semi-quantitative, even at the level of the single active electron (SAE) approximation.

At a simplest level, the classical-trajectory Monte Carlo (CTMC) method can be used to simulate the HHG process \cite{Botheron:pra09,Soifer:prl10, Higuet:pra11,Abanador:jpb17}. An extension of this method to include the phase information in the spirit of Feynman path-integral approach was developed in the quantum trajectory Monte Carlo (QTMC) model \cite{Wang:pra21,Wang:OptExp24}. The QTMC model was proposed earlier for calculating photoelectron momentum distributions (PMDs) \cite{MinLi:prl14}, and it is closely related to the semiclassical two-step model \cite{Shvetsov:pra16, Shvetsov:LasPhys25}. A modification of the SFA, the Coulomb-corrected SFA (CCSFA) \cite{Yan:prl10,Popruzhenko:pra08} and the Coulomb quantum-orbit SFA \cite{Lai2015,Faria:PhysRep20} were also proposed to take into account the effect of the Coulomb field in PMDs. Other related works using the semiclassical approach includes \cite{Hostetter:pra10,Mauger:pra16,Koch:AnnalsPhys21}.   

Difficulties in obtaining more accurate results using the semiclassical approach in many cases can be associated with the use of the Van Vleck--Gutzwiller (VVG) propagator ~\cite{Sand:prl99,Sand:pra00,Lai2015,Brennecke2020}. In fact, the drawbacks of the VVG have been well documented. They include the root search problem, the singularities at caustics, and the needs to calculate the Morse (or Maslov) indices \cite{Kay1994_1,Tannor:AnnuRev00,Brennecke2020}. To simplify the problem, the pre-exponential factor in the VVG propagator, and therefore, the Morse indices, are usually neglected. This leads to significant inaccuracies in the PMDs, as reported by Brennecke {\it et al} \cite{Brennecke2020}. With the VVG method, they managed to solve the root search problem by combining the shooting method with a clustering algorithm. They showed on an example of helium atom that very accurate two-dimensional (2D) PMD can be obtained. Unfortunately, their demonstration has been limitted to this simple target so far. The difficulties associated with the VVG propagator can largely be overcomed by replacing it with the Herman--Kluk (HK) propagator \cite{Herman1984,Kay1994_1,Grossmann:PhysLett98}.  

Recently we proposed the semiclassical strong-field Herman--Kluk propagator method (SFHK) \cite{tran2024quantum} for calculating three-dimensional (3D) PMDs of atoms and molecules in intense laser fields. Our method uses a combination of the SFA (to calculate electron wave-packet ``born" in the continuum) with the HK propagator (which accurately propagates each newly born wave-packet in the full potential). We have demonstrated that, indeed, the SFHK accurately reproduced PMDs obtained from the TDSE for various targets \cite{tran2024quantum,wcl3-x52t,mcmanus2025delay}, in a broad energy range, including the rescattering region. The purpose of this paper is to extend the SFHK to HHG calculations. We will demonstrate below that the SFHK can provide very accurate HHG yield and phase, nearly indistinguishable from the exact numerical solutions of the TDSE.  

We mention that the first attempt in HHG calculation using the HK propagator was reported in Refs.~\cite{Sand:prl99,Sand:pra00} for a one-dimensional (1D) model atom. However, it was not a true HHG process, as it did not include the ionization step. In fact, a free wave-packet was prepared in the continuum at a large distance from the atom. The same model was later used in Refs.~\cite{Zagoya:pra12,Zagoya:njp12}, with additional approximations. In similar approach was used in Zagoya {\it et al} \cite{Zagoya:njp14}, but the trajectories are not independent. Other attempts using the HK propagator for strong-field ionization treatments include Refs.~\cite{Spanner:prl03,Walser:jpb03,Xie:pra23}. We also mention the quantitative rescattering theory (QRS) \cite{Le:pra09,Lin:jpb10,Lin:book2018}, which could be classified as an extension of the three-step model, but using exact laser-free photo-recombination dipoles (for HHG) or electron-target ion elastic scattering differential cross section (for PMD). 

The rest of this paper is organized as follows. In Sec.~II.1 we describe the key elements of our SFHK method for HHG. The numerical solution of the TDSE and various computational details are given in Sec.~II.2. In Sec.~III, we demonstate the accuracies of our method on the example of HHG from atomic hydrogen and argon for different laser parameters. Here, we also diccuss the possibility of using the ADK tunneling theory \cite{ADK:JETP86} to decscribe the ionization step in the SFHK. Finally we finish our paper with a summary and outlook. We also provide extensive details of our theory in the Appendices.

\section{Theoretical Methods}
\subsection{The strong-field Herman-Kluk (SFHK) method for HHG}  

The SFHK method has been proposed and described in Ref.~\cite{tran2024quantum} for calculating PMDs of atoms and molecules in intense laser fields. More details about the method can also be found in Refs.~\cite{wcl3-x52t,mcmanus2025delay}. In all previous applications of the SFHK, the PMDs were found to be in very good agreements with exact numerical solutions of the TDSE for different targets. The origin of these high accuracies can be attributed to the accuracies of the HK propagator which includes the full account of the atomic (or molecular) potential together with the laser interaction right after the tunneling ionization step. But it also implies rather high accuracies of the SFA for describing the continuum electron wave-packet immediately after ionization. Below we only describe the key points in our method, as it is extended to HHG calculations. More detailed information is also provided in Appendices A-D. 

The main task in HHG calculations is to calculate time-dependent induced dipole $D_z(t) = - \left\langle\ \Psi(t) \right| z \left| \Psi(t) \right\rangle$. Here and in the following, atomic units will be used throughout, unless otherwise indicated. Equivalently, one can also calculate the dipole velocity and/or dipole acceleration \cite{Burnett:pra92,Le:pra09,BandraukPRA2009}. Here, the laser is assumed to be linearly polarized along $z$-axis and the state vector $|\Psi(t) \rangle$ is typically assumed to evolve in time in the laser field starting from the initial ground state $|\psi_0(t=0) \rangle$. At the microscopic level, the HHG yield is proportional to the Fourier transform of the induced dipole. It can also be expressed through the Fourier transform of the dipole velocity or dipole acceleration. To be more specific, we focus below on the calculation of the induced dipole $D_z(t)$, although in the SFHK we typically calculate both the dipole and dipole velocity. 

We can formally expand our state vector as
\begin{equation}
\left| \Psi(t) \right\rangle = a\left| \psi_0(t) \right\rangle + \left| \psi_e(t) \right\rangle + \left| \psi_c(t) \right\rangle
\label{WF-expansion}
\end{equation} 

Here, $|\psi_e(t)\rangle$ describes the excited bound states and $|\psi_c(t) \rangle$ describes the electron ionized to the continuum. Parameter $a$ describes the depletion of the ground state. For sufficiently low laser intensities one can make an approximation $a \approx 1$. Note that in typical experiments, it is desirable to keep the laser intensity rather low, with the probability of ionization less than only a few percents, such that the phase matching condition for HHG can be optimized \cite{Gaarde:jpb08,Popmintchev:science12}.

The excited states might play some roles in the ionization. However, for the HHG process under typical conditions, the excited states have been found to have minor effects on the overall shape of the spectra above the threshold. Under the above assumptions, the expression for the induced dipole reduces to 
\begin{equation}
\begin{split}
D_z(t) & = - \left\langle\ \Psi(t) \right| z \left| \Psi(t) \right\rangle \\
& \approx - \left\langle\ \psi_0(t) + \psi_c(t) \right| z \left| \psi_0(t) + \psi_c(t) \right\rangle \\
& \approx - \left\langle\ \psi_0e^{iI_pt} \right| z \left| \psi_c(t) \right\rangle + c.c. \\ 
\end{split}
\label{induced dipole}
\end{equation} 

Our task is therefore reduces to the calculation of the continuum part of the wavefunction. We remark that in the above equation, for simplicity the continuum-continuum transition $\left\langle \psi_c(t)|z|\psi_c(t) \right\rangle$ is neglected, as its contribution is expected to be much smaller than the recombination from the continuum back to the ground state. In principle, the continuum-continuum contribution can also be included, once we know the continuum wavefunction.  

The essence of the SFHK method can now be described as follows. First, at any given time $t_0$ during the laser pulse, we consider a newly created continuum part of the wavefunction $\Delta\psi_c({\bf{p}},t)$ that was ``born" in the time window $\Delta t$ just preceding $t_0$. Here, for convenience we work with the wavefunction in the momentum space. This part of the wavefunction can be quite accurately modeled by the SFA \cite{Ivanov:jmo05,Bauer:pra05,Milosevic:jpb06,Lai2015,Brennecke2019,Brennecke2020}. Note that, in contrast to the ``standard" SFA, we only use the saddle-point approximation (SPA) to get this part of the wavefunction right at the tunnel exit when it was just born to the continuum -- see Appendix A. Second, the subsequent propagation of the electron in the continuum in the combined potential of the atomic potential $V_a$ and the interaction with the laser electric field $V_L(t)$ is then governed by the exact quantum propagator $K({\bf{p}},t;{\bf{k}},t_0)$, which we approximate by the semiclassical Herman-Kluk propagator \cite{Herman1984, Kluk1986, Kay1994_1, Kay1994_2, Kay1994_3}, namely
\begin{equation}
\begin{split}
\Delta\psi_c({\bf{p}},t) & = \int d{\bf{k}} K({\bf{p}},t;{\bf{k}},t_0) \Delta\psi_c({\bf{k}},t_0) \\
& \approx \int d{\bf{k}} K^{HK}({\bf{p}},t;{\bf{k}},t_0) \Delta\psi_c({\bf{k}},t_0)
\end{split}
\label{propagation}
\end{equation} 

And third, we add contributions originated from all ionization times during the laser pulse. The final result is then given by (see Appendix A for more detail)
\begin{equation}
\begin{split}
\psi_c({\bf{p}},t) &  \propto \sum_s \iint d{\bf{p}}_{t_0} d{\bf{q}}_{t_0} \braket{{\bf{p}}|{\bf{p}}_t,{\bf{q}}_t,\gamma} C({\bf{p}}_{t_0},{\bf{q}}_{t_0},t) \\
&\times \exp{\left[ i(S_\to + S_\downarrow^0 + {\bf{p}}_{t_0} \cdot {\bf{q}}_{t_0}) \right]} \\
&\times \dfrac{\left\langle\ {\bf{p}}_{t_0} + {\bf{A}}(t_s) -  {\bf{A}}(t_0) \right| {\bf{r}} \cdot {\bf{E}}(t_s) \left| \psi_0 \right\rangle}{\sqrt{{\bf{E}}(t_s)\cdot [{\bf{p}}_{t_0} + {\bf{A}}(t_s) -  {\bf{A}}(t_0)] }}.
\end{split}
\label{SFHK-wf}
\end{equation} 
Here ${\bf{E}}$ and ${\bf{A}}$ are the laser electric field and vector potential, respectively, and $\braket{{\bf{p}}|{\bf{p}}_t,{\bf{q}}_t,\gamma}$ is the coherent state basis function in the momentum representation at time $t$, with the center at $({\bf{p}}_t,{\bf{q}}_t)$ in the phase space. The evolution of each trajectory, indicated by $({\bf{p}}_t,{\bf{q}}_t)$, is governed by classical Hamilton's equations of motion. For more detail and other notations, see Appendix A. 

We emphasize again that the full interaction is included in this treatment at all time, starting right after the tunneling ionization step at the tunnel exit, and there are no further approximations made to the HK propagator. In particular, the pre-exponent factor $C({\bf{p}}_{t_0},{\bf{q}}_{t_0},t)$ is fully included. Within the SFHK, there is no need to calculate the Morse index explicitly, as in typical methods based on the VVG propagator \cite{Brennecke2020}. 

Note that due to the restrictions imposed by the SPA equation, the integration over the initial momentum after tunneling is effectively only two-dimensional. Furthermore, we also impose the condition ${\bf{q}}_{t_0} = {\rm Re}[\int_{t_s}^{t_0}{\bf{A}}(\tau) d\tau]$ at the tunnel exit \cite{Yan:prl2010,Lai2015,Brennecke2020}. In practical implementations of Eq.~(\ref{SFHK-wf}), the integral over the initial distributions can be carried out conveniently using standard Monte-Carlo methods, as typically done in semiclassical trajectory-based methods. 

To further simplify the calculations, we expand the initial wavefunction $\psi_0$ in Gaussian-type orbitals (GTOs) \cite{Hehre:jcp69,Ditchfield:jcp71}, as commonly done in standard quantum chemistry software such as {\em Gaussian} and {\em Gamess} \cite{g09,GAMESS}. For completeness, the calculations of typical dipole matrix elements used in the SFHK are described in detail in Appendices B and C. 

We remark that for atomic hydrogen, in principle we can use the modified SPA (see, for example, review by Milosevic {\it et al} \cite{Milosevic:jpb06}) instead of the standard SPA. This leads to a slightly different equation for the total wavefunction in the continuum -- see Appendix B.2. In practice, we found that the direct use of Eq.~(4) together with approximate expansions of the hydrogen ground state wavefunction in the GTO basis lead to nearly identical results for HHG as compared to the ones obtained with the modified SPA.

\subsection{Numerical solution of the time-dependent Schr\"odinger equation for HHG}

In this subsection, we provide the theoretical formalism for the integration of the non-relativistic time-dependent Schr\"odinger equation (TDSE) in the single-active-electron (SAE) approximation we have developed for polyatomic molecules~\cite{GoetzMolecules},  which we briefly describe here and apply to the specific case of an atomic target.  The suite of codes runs on distributed and shared memory parallel computers and can be used to study strong-field ionization and high-order harmonic generation in polyatomic molecules in different gauges and dipole forms. 

In the electric dipole approximation, the non-relativistic time-dependent Schr\"odinger equation reads (atomic units used throught),
\begin{eqnarray}
\label{eq:TDSE1}
    i\dfrac{\partial}{\partial t}|\Psi(t)\rangle = \big[ \hat{H}_0 + \hat{V}_L(t) + \hat{W}_{C} \big] |\Psi(t)\rangle\,,
\end{eqnarray}
where $\hat{H}_0 = -\dfrac{1}{2}\boldsymbol{\nabla}^2 + \hat{V}_a(\boldsymbol{r})$ describes the field-free Hamiltonian with $\boldsymbol{\nabla}^2$ the Laplacian and $V_a(\boldsymbol{r})$ a model for the potential energy interaction between the single-active electron and the parent ion. The term $\hat{V}_L(t) = -\boldsymbol{\mu}\cdot \boldsymbol{E}(t)$ describes the laser-matter interaction in the non-relativistic electric dipole approximation and Length Gauge (LG) with $\boldsymbol{\mu} = -\boldsymbol{r}$ the dipole moment. $\hat{W}_C = -i\,\gamma_C(r)$ is an complex absorbing potential (CAP) introduced to avoid reflections off the grid boundaries during time-propagation. Following Ref.~\cite{GreenmanPRA2010}, $\gamma_C(r) = \eta_C\,\Theta(r-r_C)\, (r-r_C)^2$, with $\eta_C$ the strength of the absorbing potential and $\Theta(r-r_C)$ the Heaviside function. 

In order to integrate Eq.\eqref{eq:TDSE1}, the  state vector is described as a linear combination of eigenfunctions $|\Phi_n\rangle$ of the field-free Hamiltonian $\hat{H}_0$, i.e., $\hat{H}_0|\Phi_n\rangle = \epsilon_n|\Phi_n\rangle$ with eigenenergies $\epsilon_n$,
\begin{eqnarray}
\label{eq:TDSE2}
    |\Psi(t)\rangle =\sum_{i=1}^{N_{E}} e^{-i\,\epsilon_n\,(t-t_o)}\,C_n(t)|\Phi_n\rangle\,,
\end{eqnarray}
 Inserting Eq.~\eqref{eq:TDSE2} into Eq.~\eqref{eq:TDSE1} yields the coupled equations of motion for the time-dependent coefficients $C_n(t)$,
\begin{eqnarray}
\label{eq:TDSE3}
    \dfrac{d}{dt}C_{n}(t)=
    i\sum_{m=1}^{N_E}\boldsymbol{E}(t)\cdot\boldsymbol{\mu}_{n,m}\, C_{m}(t) + 
    \sum_{m=1}^{N_E} \gamma_{n,m}\, C_{m}(t),\,\,
\end{eqnarray}
where $\boldsymbol{\mu_{n,m}} = -\langle \Phi_n|\boldsymbol{r}|\Phi_m\rangle$ designates the transition dipole elements in the basis of eigenfunctions of $\hat{H}_0$, and $\gamma_{n,m} = \langle\Psi_n|\gamma_C(r)|\Phi_m\rangle$. Eq.~\eqref{eq:TDSE3} is solved using a fourth-order explicit Runge-Kutta scheme~\cite{BUTCHER1996247} with initial condition $|\Phi(t_0)\rangle$ corresponding ground electronic state of the target system. 

To obtain the field-free eigenstates efficiently, we employ a finite-element discrete variable representation\cite{RescignoPRA2000} for the radial part of the eigenfunction and express the angular part as a linear combination of tesseral spherical harmonics, 
\begin{eqnarray}
    \label{eq:TISE1}
        \Phi_n(\boldsymbol{r}_i) = \sum_{l=0}^{L_{max}}\sum_{m=-l}^{+l}
    \dfrac{b^{n}_{l,m}(r_i)}{r_i \sqrt{w_i}} T_{l,m}(\theta,\phi)\,,
\end{eqnarray}
with $T_{l,m}$ the tesseral spherical harmonics~\cite{VMK}and $w_i$ the Gaussian weight associated to the radial grid-point $r_i$. For molecular systems, the couplings between the different basis with different $l$ and $m$ due to the molecular potential $V_a(\boldsymbol{r})$ makes diagnonalization computationally demanding, if not untracktable. In the finite-element basis, however, the Hamiltonian $H_0$ is sparse. We take advantange of such sparsity and obtain the eigenvalues of $\hat{H}_0$ using
the FEAST library package~\cite{FEAST} by splitting the energy domain of the desired eigenfunctions into $N$ subdomains which are diagonalized independently. For atomic systems centred at the origin and subject to linearly polarized electric field the summation over $m$ in Eq.~\eqref{eq:TISE1} reduces to $m=0$ and diagonalization can be carried independently for each angular momentum $l$ due to the spherically symmetric nature of the atomic potential.

The electric field $E(t)$ is parametrized according to,

\begin{eqnarray}
\label{EfieldTDSE}
    \boldsymbol{E}(t) = \sum_{i=1}^{3} E_i\, h_i(t-t_i)\,\cos(\omega_i(t-t_i)+\phi_i)\,\boldsymbol{\hat{e}}_i
\end{eqnarray}
with $\boldsymbol{\hat{e}}_i$ the cartesian unit vectors in the laboratory frame of reference, $E_i$ the peak-field amplitude, $\omega_i$ the central circular frequency, $\phi_i$ the carrier-envelope phase (CEP), $t_i$ the time delay and $h_i(t-t_i)$ a envelope function. 

\emph{Observables ---} Upon obtaining the state vector~\eqref{eq:TDSE2} the HHG spectrum can be obtained from Fourier transform of the time-dependent expectation values
of the dipole, dipole velocity, and dipole acceleration~\cite{HanPRA2010}. In the LG, the dipole acceleration reads,
\begin{subequations}
\label{eqn:TDSE_forms}
\begin{eqnarray}
    \label{eqn:dipa}
    \langle\boldsymbol{a}(t)\rangle = -\langle\Psi(t)|\boldsymbol{\nabla} \hat{U}_M(\boldsymbol{r})|\Psi(t)\rangle\,,
\end{eqnarray}
with $\boldsymbol{\nabla} \hat{U}_M(\boldsymbol{r}) = \boldsymbol{\nabla} \hat{V}_a(\boldsymbol{r})- \boldsymbol{E}(t)$ the gradient of the combined  molecular (atomic) potential energy and laser-induced dipole interaction. Alternatively, the HHG spectrum can be computed from the dipole velocity, which in the LG is given by,
\begin{eqnarray}
\label{eqn:dipv}
    \langle\boldsymbol{p}(t)\rangle = \langle\Psi(t)|\boldsymbol{p}|\Psi(t)\rangle\,,
\end{eqnarray}
with $\boldsymbol{p}=-i\boldsymbol{\nabla}$ the momentum operator, and finally the expectation value of field-induced dipole moment,
\begin{eqnarray}
    \label{eqn:dipm}
    \langle\boldsymbol{D}(t)\rangle = -\langle\Psi(t)|\boldsymbol{r}|\Psi(t)\rangle\,,
\end{eqnarray}
In agreement with Ref.~\cite{BandraukPRA2009}, have confirmed that all three dipole forms~\eqref{eqn:dipa},\eqref{eqn:dipv} and~\eqref{eqn:dipm} are related by,
\begin{eqnarray}
    |\boldsymbol{a}(\omega)|^2\approx |\boldsymbol{p}(\omega)|^2\,\omega^2\approx |\boldsymbol{D}(\omega)|^2\,\omega^4\,,
\end{eqnarray}
\end{subequations}
with $\boldsymbol{a}(\omega)$, $\boldsymbol{p}(\omega)$ and $\boldsymbol{D}(\omega)$ the Fourier transform the dipole forms in Eqs.~\eqref{eqn:dipa},~\eqref{eqn:dipv} and~\eqref{eqn:dipm}, respectively,
for weak electric field intensities resulting in small ionization probabilities. We also used the Hanning window function \cite{Camp:jpb18} in the Fourier transform to drop the dipole smoothly to zero when laser ends  -- see Appendix D. The latter is defined by the lost of norm due to the CAP.

For the simulations reported in this work,
the electric field 
is asummed to be linearly polarized in the $\boldsymbol{\hat{e}}_z$ direction with maximum peak intensity $I_0=[1 , 1.2]\times 10^{14}\mathrm{Wcm}^{-2}$, duration of few cycles and flat spectral phase. The field envelope function is given by $h_i(t-ti)=\sin^2((t-t_a)/(t_b-t_a))$ with $t_a=t_i-\Delta T$ and $t_b=t_i+\Delta T$, with $\Delta T$ the pulse duration, if $t_a\le t\le t_b$ and $h_i(t-ti)=0$ otherwise.  After the electric field is over, the time propagation is continued for 5 additional cycles to ensure slow components of the photoionized electron reach the CAP. The CAP strength is set to $\gamma_C = 0.005\,$a.u. and $r_C = 270\,$ bohr. For the field-free eigenstates, the radial grid is set to $r_{max}=300\,$ bohr with $N_{\mathrm{DVR}} = 1000$ mapped Legendre-Gauss-Lobatto collocation points for the radial coordinate and $L_{max}=300$ with fixed $m=0$. Finally, the summations over eigenfunctions of $\hat{H}_0$ in Eq.~\eqref{eq:TDSE3} are truncanted to $N_E$, corresponding to an energy cutoff of $\epsilon_{N_E}=20$ a.u., ensuring convergence of the HHG spectrum with respect to $N_E$. Equation~\eqref{eq:TDSE3} is integrated using a four-order explicit Runge-Kutta scheme with a time step of $\Delta t = 0.02\,$ a.u.

\section{Results and Discussions}
\subsection{HHG from atomic hydrogen} 
In this Subsection, we will focus on HHG from atomic hydrogen at different laser parameters and analyze some features revealed by our SFHK method.

\begin{figure}[tb!]
\begin{center}
\includegraphics[width=0.95\linewidth]{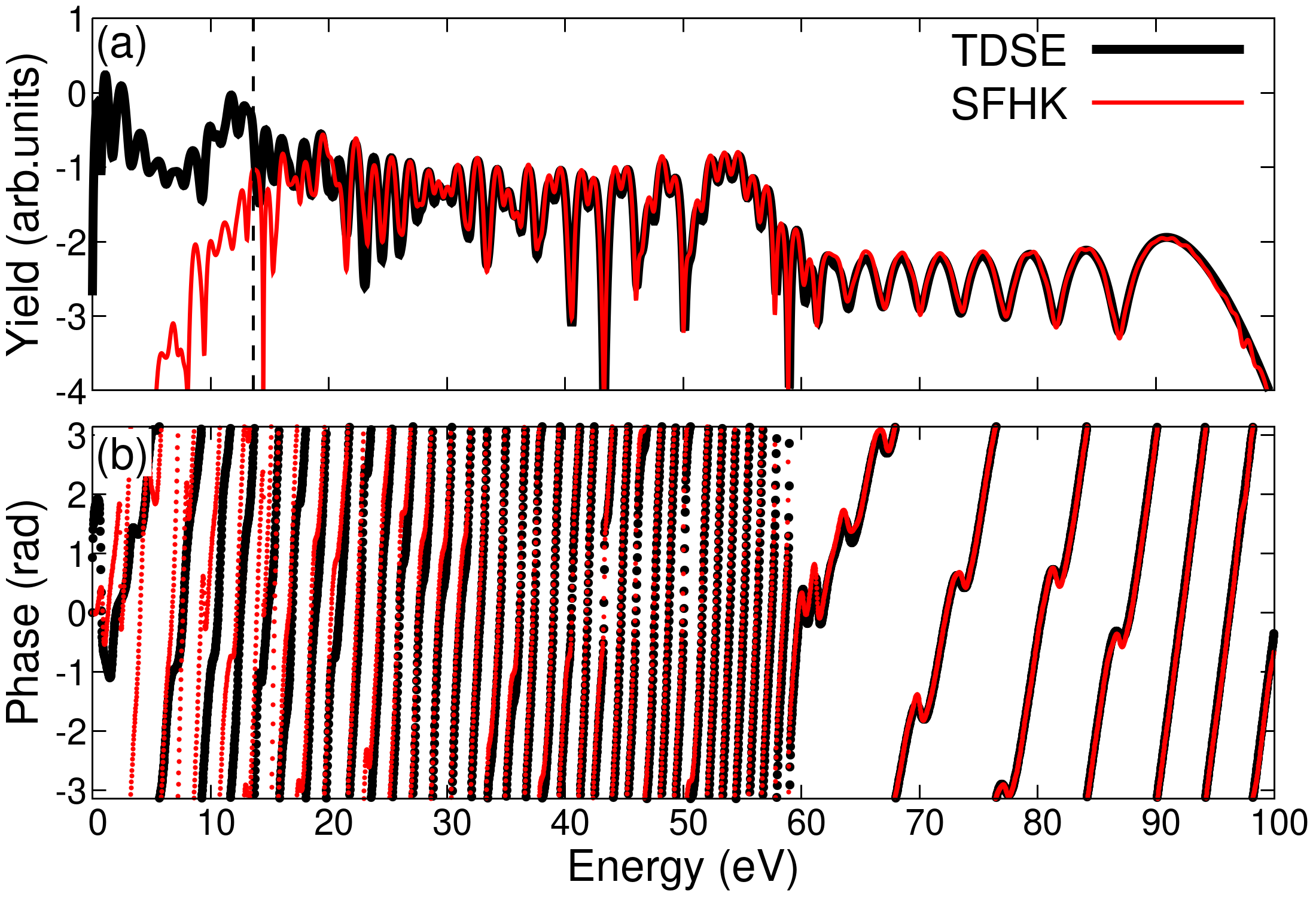}
\caption{Comparison of HHG yield (a) and phase (b), vs emitted photon energy, calculated within the SFHK and TDSE for atomic hydrogen. The vertical axis in (a) is on a logarithmic scale. A three-cycle laser pulse with the wavelength of 1,600-nm, intensity of $10^{14}$ W/cm$^2$, and CEP $\phi=0$ was used in the calculations.}
\label{fig:H1600}
\end{center}
\end{figure}

\begin{figure}[tb!]
\begin{center}
\includegraphics[width=0.95\linewidth]{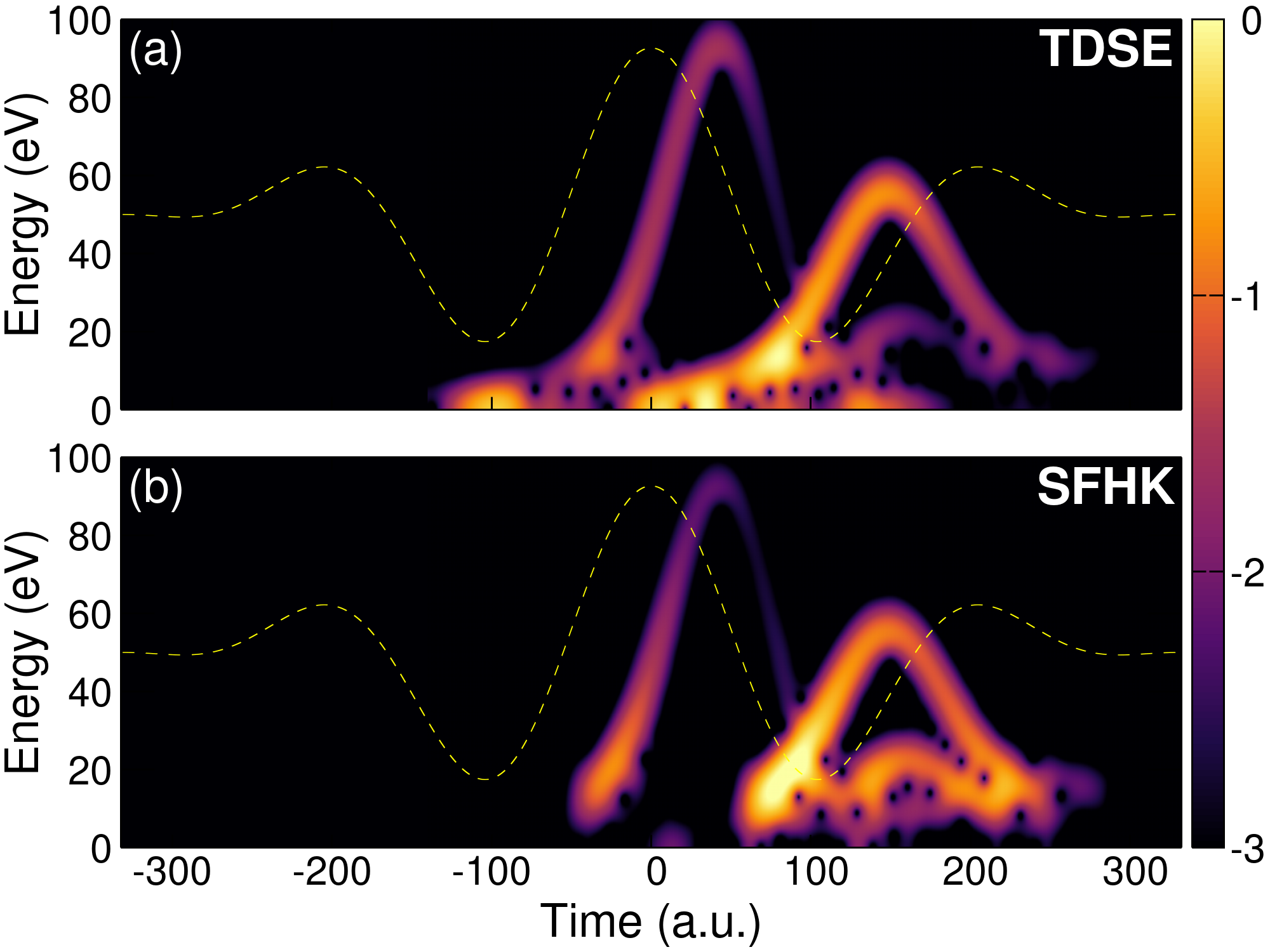}
\caption{The Gabor time-frequency analysis of the induced dipole velocity calculated with the TDSE (a) and the SFHK (b). The color map is on a logarithmic scale. The electric field of the laser pulse is also shown as the dashed line. The laser parameters are the same as in Fig.~\ref{fig:H1600}.}
\label{fig:HGabor}
\end{center}
\end{figure}

First in Fig.~\ref{fig:H1600}(a), we compare the HHG spectrum obtained by the TDSE and the SFHK for a three-cycle laser pulse with a wavelength of 1,600-nm, intensity of $10^{14}$ W/cm$^2$, and carrier-envelope phase (CEP) of 0 degrees. The agreement between the two methods is excellent in the whole range of harmonic energy slightly above the threshold (at 13.6 eV, indicated by the vertical dashed line in the figure). Furthermore, the HHG dipole phases calculated from the two methods also agree well in the same energy range; see Fig.~\ref{fig:H1600}(b). To the best of our knowledge, this level of accuracy has never been demonstrated for any approximate semiclassical methods. Clearly, the SFHK result would be very inaccurate if the Coulomb interaction was neglected. In fact, if the atomic potential (the Coulomb potential in this case) is neglected, then our method reproduces the Lewenstein model \cite{Lewenstein:pra94}. The origin of the step-like structure in the HHG yields near 55 eV and the corresponding rapid changes in the harmonic dipole phase will be discussed below in the analysis of the Gabor transform in Fig.~\ref{fig:HGabor}.

The discrepancies below the threshold can be easily understood. This happens because within the SFHK we totally neglect the effect of the bound excited states during the ionization and recombination; see the discussion in Sec.~II.1. Furthermore, we also neglect the effect of continuum-continuum transition. The excellent agreement between the two methods indicates that these effects are not noticeable for energies more than a few eVs above the threshold.                

We note that the SFHK result in Fig.~\ref{fig:H1600} was obtained within the velocity form of the dipole. The use of the induced dipole itself leads to nearly indistinguishable results (not shown) for both yield and phase, as compared to that of the velocity form. Furthermore, we have used the modified SPA (see, for example, \cite{Milosevic:jpb06} as well as Appendix B.2) for the SFA in the ionization step. We found that quite accurate results can also be obtained if the standard SPA is used together with the initial ground-state wavefunction approximated by a combination of the GTOs. This implies that for practical purposes one can use the wavefunctions expressed through the GTO basis sets that can conveniently be calculated by standard quantum chemistry software such as {\em Gaussian} or {\em Gamess} for any atomic or molecular targets.  

To provide further benchmarks for the SFHK, we compare in Fig.~\ref{fig:HGabor} the Gabor time-frequency analysis of the induced dipole velocity calculated with the TDSE (in the upper panel) and the SFHK (in the lower panel). Again, the two methods agree well, except for the energies near and below the threshold. As shown in Fig.~~\ref{fig:HGabor}, due to the shortness of the laser pulse, most of the harmonics were generated by two to four ``bursts" only. As can be seen from the figure, for energies above 60 eV, there are only two contributing pathways, the so-called short and long trajectories, for the emission time $t\lessapprox 40$ a.u. and $t\gtrapprox 40$ a.u., respectively. The interference between these two contributions gives rise to the slow oscillations in the HHG yields above 60 eV observed in Fig.~\ref{fig:H1600}(a). Below 60 eV, two additional contributions can be seen. These additional pathways are from electrons ionized near the peak of the pulse. Therefore, their signal is significantly stronger. This leads to the enhanced step-like structure in the HHG yields near 55 eV in Fig.~\ref{fig:H1600}(a). Interference of all these four contributions gives rise to much faster oscillation structures below 55 eV as compared to the plateau above 60 eV. Correspondingly, the dipole phase below 60 eV changes more rapidly as compared to that from above 60 eV, as can be seen in Fig.~\ref{fig:H1600}(b).   

\begin{figure}[tb!]
\begin{center}
\includegraphics[width=0.95\linewidth]{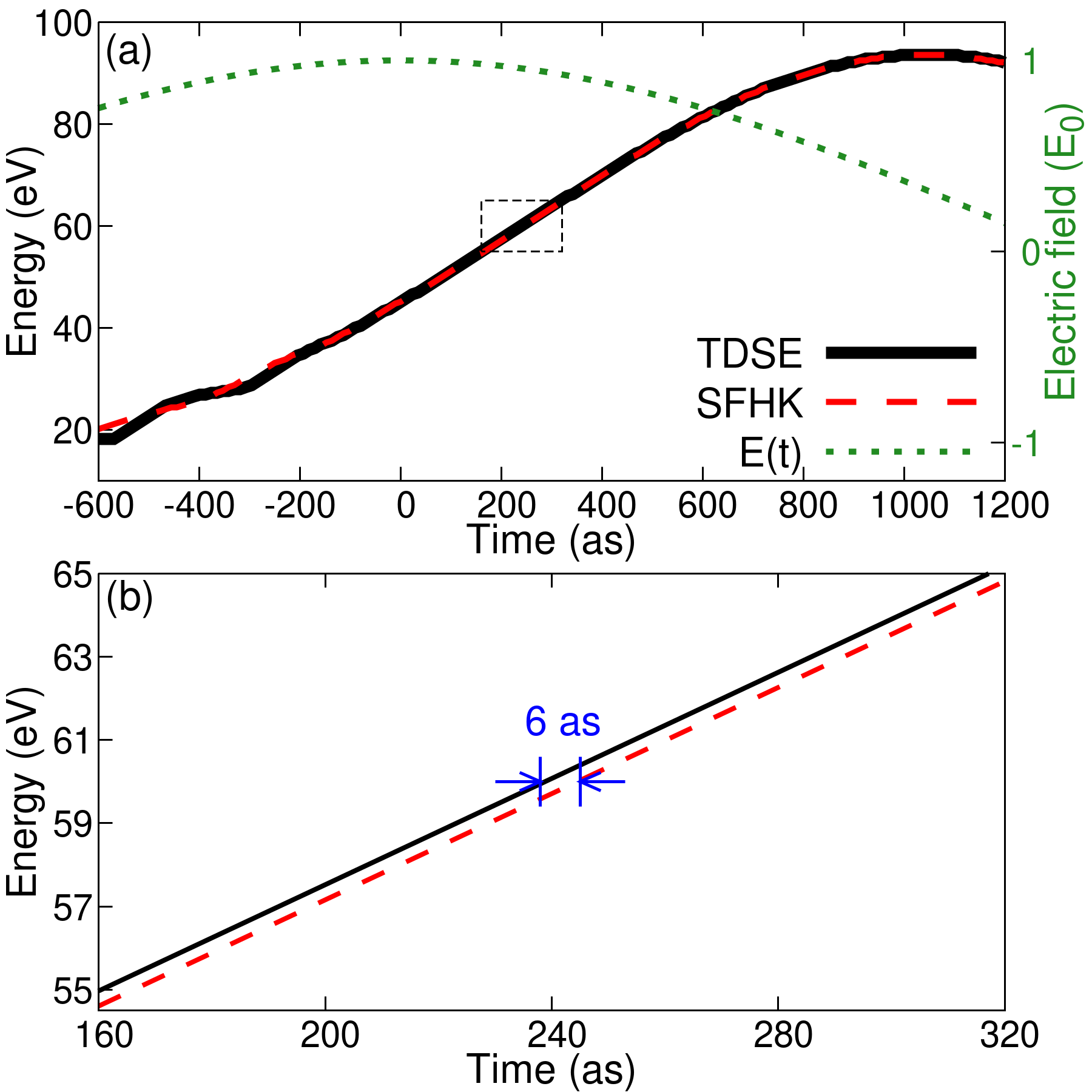}
\caption{(a) Comparison of the emission time calculated using the Gabor transform of the dipole velocity from the SFHK and the TDSE. See text for more detail. The electric field of the laser pulse is also shown as the dotted line. (b) The zoom-in of (a) in the time range between 160 as to 320 as. The laser parameters are the same as in Fig.~\ref{fig:H1600}.}
\label{fig:HGabortime}
\end{center}
\end{figure}

To have a more quantitative comparison, we now analyze the first burst that occurred near the peak of the pulse (near time $t=0$). For that purpose, we calculated for any given harmonic energy the center of the first burst as a function of time and the result is shown in Fig.~\ref{fig:HGabortime}(a). Note that in HHG spectroscopy, the center of this burst has been interpreted as the recombination time. As the two methods, TDSE and SFHK, lead to nearly the same results, we also show in Fig.~\ref{fig:HGabortime}(b) the zoom-in comparison. In fact, the SFHK gives a recombination time delayed by about 6 attoseconds as compared to the TDSE result.       

\begin{figure}[tb!]
\begin{center}
\includegraphics[width=0.95\linewidth]{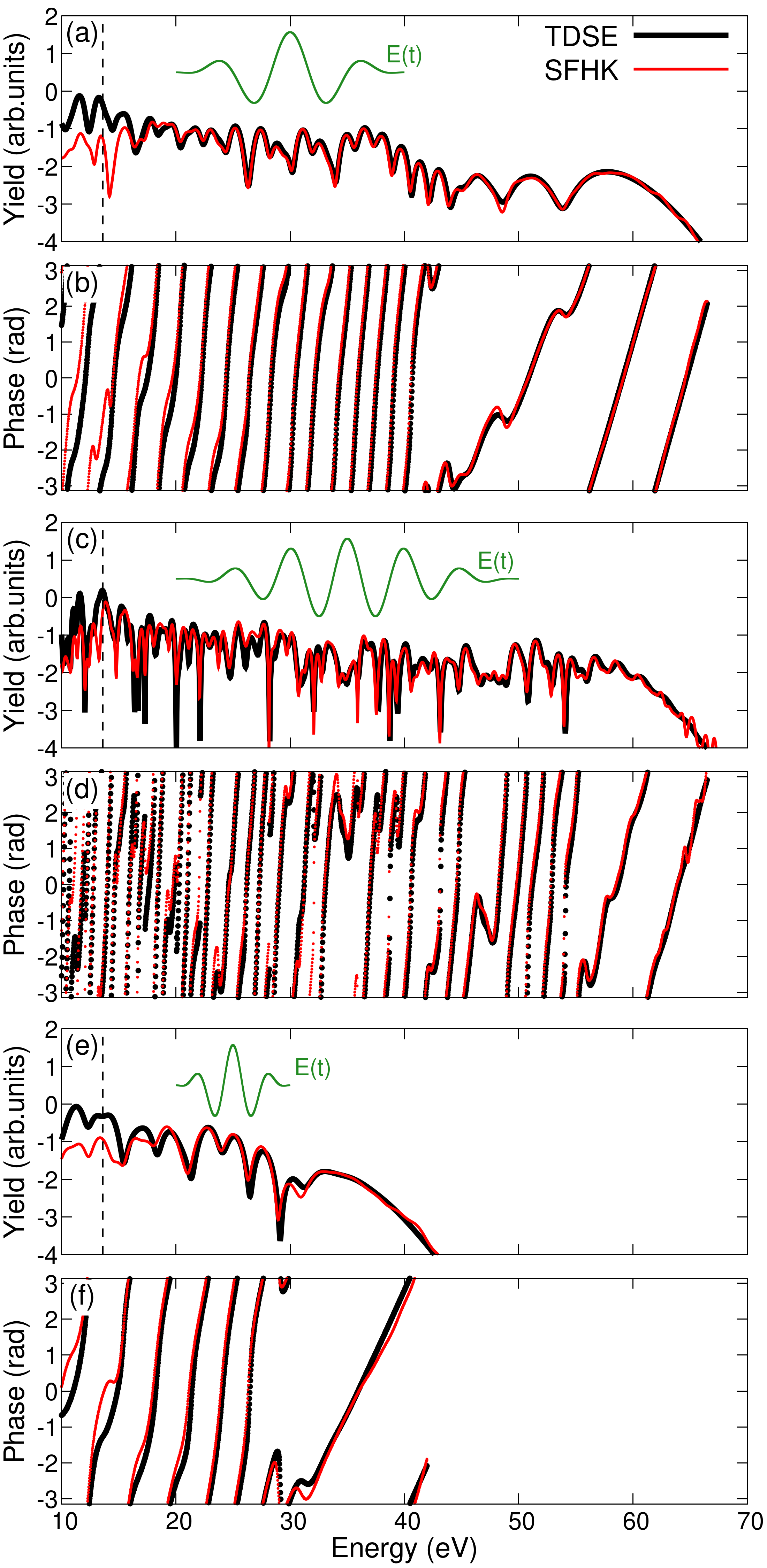}
\caption{(a) and (b): same as in Fig.~\ref{fig:H1600} but for the wavelength of 1,200-nm. (c) and (d): same as in (a) and (b), respectively, but for a six-cycle laser pulse. (e) and (f): same as (a) and (b) but for the wavelength of 800-nm. The vertical axis in (a), (c), and (e) is on a logarithmic scale.}
\label{fig:H1200}
\end{center}
\end{figure}

Next, we compare our SFHK results with the TDSE for the case of the wavelength of 1,200-nm in Fig.~\ref{fig:H1200}(a,b) and (c,d) for the three-cycle and six-cycle pulse, respectively. The laser intensity remains the same as in Fig.~\ref{fig:H1600} at $10^{14}$ W/cm$^2$. We can see similar excellent agreements between the two methods in the HHG yields and dipole phases for both cases. The SFHK results deteriorate a bit as the wavelength decreases to 800-nm, but they are still quite good -- see Fig.~\ref{fig:H1200}(e,f). In all these SFHK calculations, the width of the coherent states was chosen to be $\gamma=0.5$ a.u. To reduce effect of highly unstable trajectories, we implement the trajectory removal technique proposed in Ref.~\cite{Kay1994_3}. Specifically, we impose a cutoff function, $D_t$, on the value of $|C({\bf{p}}_{t_0},{\bf{q}}_{t_0},t)|^2$ in Eq.~(\ref{SFHK-wf}) such that any trajectory having $|C({\bf{p}}_{t_0},{\bf{q}}_{t_0},t)|^2 > D_t$ is removed. In this work, we chose $D_t = 10^{30}$. It is noticed that the exponent in the value of $D_t$ is related to Lyapunov numbers \cite{Kay1994_3}. We have checked that our results are quite stable with respect to these parameters in the range of $\gamma=[0.1:2]$ a.u. and $D_t =[10^{20}:10^{35}]$. To get converged SFHK results we typically used about 2 million to 20 million trajectories.        

\begin{figure}[tb!]
\begin{center}
\includegraphics[width=0.95\linewidth]{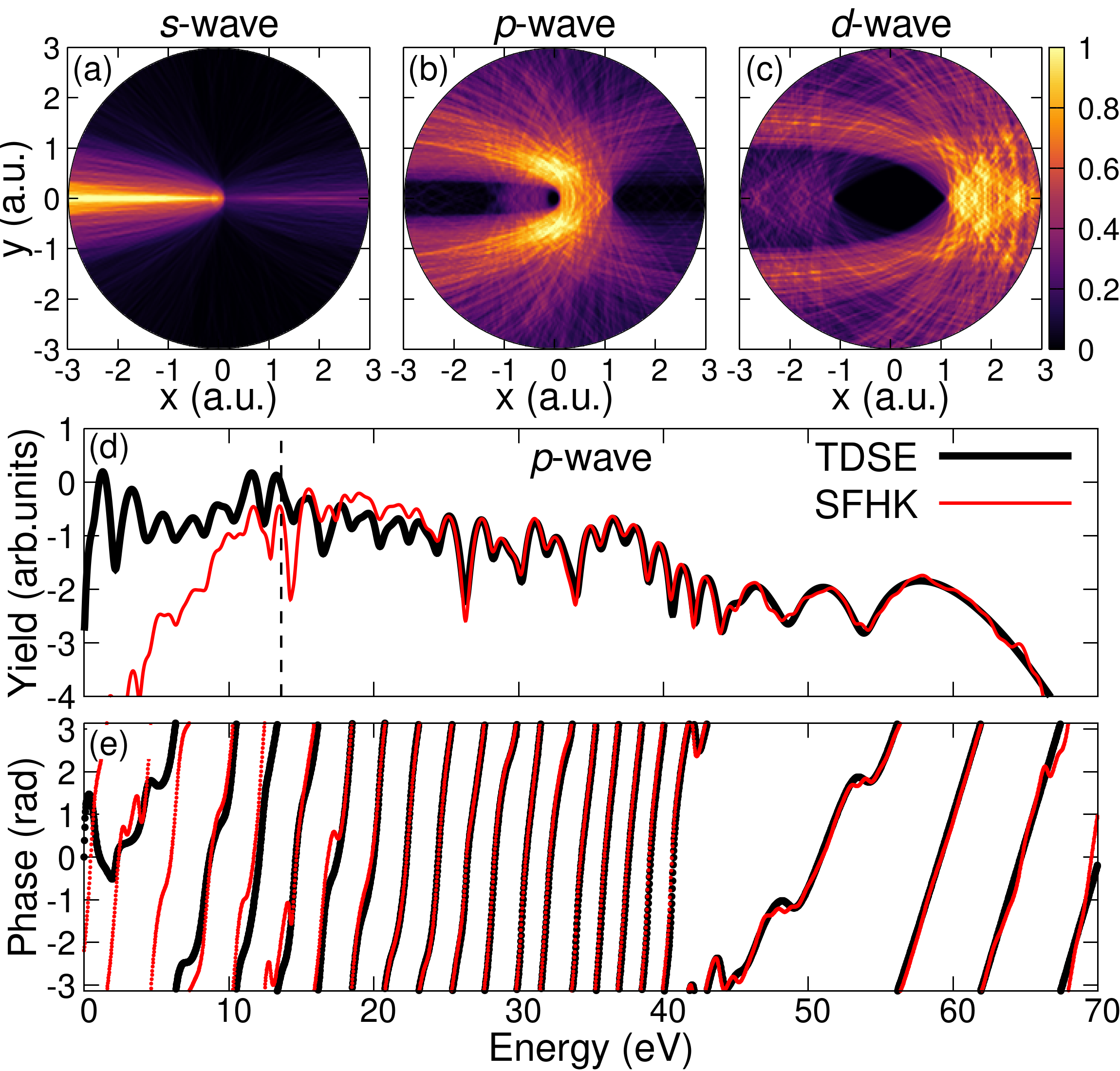}
\caption{(a), (b), and (c): Partial-wave analysis of the trajectories returning to within $r < 3$ a.u. from the core, for $s$-wave, $p$-wave, and $d$-wave, respectively, for atomic hydrogen case. See text for more detail. (d) and (e): comparison of HHG yield and phase calculated with the TDSE with the SFHK, but only $p$-wave trajectories that come close to within $r < 3$ a.u. from the core were included in the SFHK. The vertical axis in (d) is on a logarithmic scale. The laser parameters are the same as in Fig.~\ref{fig:H1200}(a) and (b).}
\label{fig:Hpwave}
\end{center}
\end{figure}

According to the three-step model as well as the QRS theory, harmonic emission occurred when the electron that was released earlier into the continuum recombines with the parent ion as it is driven back close to the core by the laser electric field. Furthermore, we can expect that when the electron is sufficiently close to the core, the Coulomb interaction dominates over the laser field, and therefore the electron angular momentum can be considered to be nearly constant in that region. Samples of those trajectories are shown in Fig.~\ref{fig:Hpwave}(a), (b), and (c) for ``$s$-wave", ``$p$-wave", and ``$d$-wave", respectively, in the region of radius $r < 3$ a.u. from the nucleus placed at the origin. Here, the laser parameters are the same as in Fig.~\ref{fig:H1200}(a), and with the linear polarization along the $x$-axis. For our purpose, we assign the label $s$-, $p$-, and $d$-wave for the trajectories, when the electron angular momentum near the core satisfies the condition $l \in [0:0.5)$, $l \in [0.5:1.5)$, $l \in [1.5:2.5)$, respectively. Note that these plots are quite asymmetric with respect to the reflection at $x=0$. This is due to the shortness of the three-cycle pulse used in this case. Due to the selection rules, one can expect that we only need to include ``$p$-wave" trajectories for the HHG calculation within the SFHK method as the hydrogen ground state is $1s$. This is indeed the case as demonstrated in Fig.~\ref{fig:Hpwave}(d) and (e). Here, we show the HHG yield and phase obtained within the SFHK, in which we include only $p$-wave trajectories that come close to the core within $r < 3$ a.u. The results are indeed in very good agreement with the TDSE. Apart from providing physical intuitions, these results indicate that the implementations of the SFHK in practice can be significantly improved if we take into account symmetry consideration as well as other possible restrictions on the classical trajectories. This will be investigated in the future.

\subsection{HHG for argon and the Cooper minimum}

In the previous subsection, we have demonstrated the accuracies of the SFHK in the calculations of HHG from atomic hydrogen. In this subsection, we want to check the SFHK method in a more challenging test. For this purpose, we choose argon as the target, as HHG spectra from argon have been shown to have a pronounced Cooper minimum near 50 eV, both theoretically \cite{Le:pra08,Le:pra09} and experimentally \cite{Worner:prl09,Higuet:pra11}. According to the QRS theory \cite{Le:pra09,Lin:jpb10,Lin:book2018}, the origin of this Cooper minimum in HHG spectra can be traced back to the Cooper minimum in the photoionization (or photo-recombination) differential cross section. Therefore, it is quite stable with respect to the laser parameters. 

\begin{figure}[tb!]
\begin{center}
\includegraphics[width=0.95\linewidth]{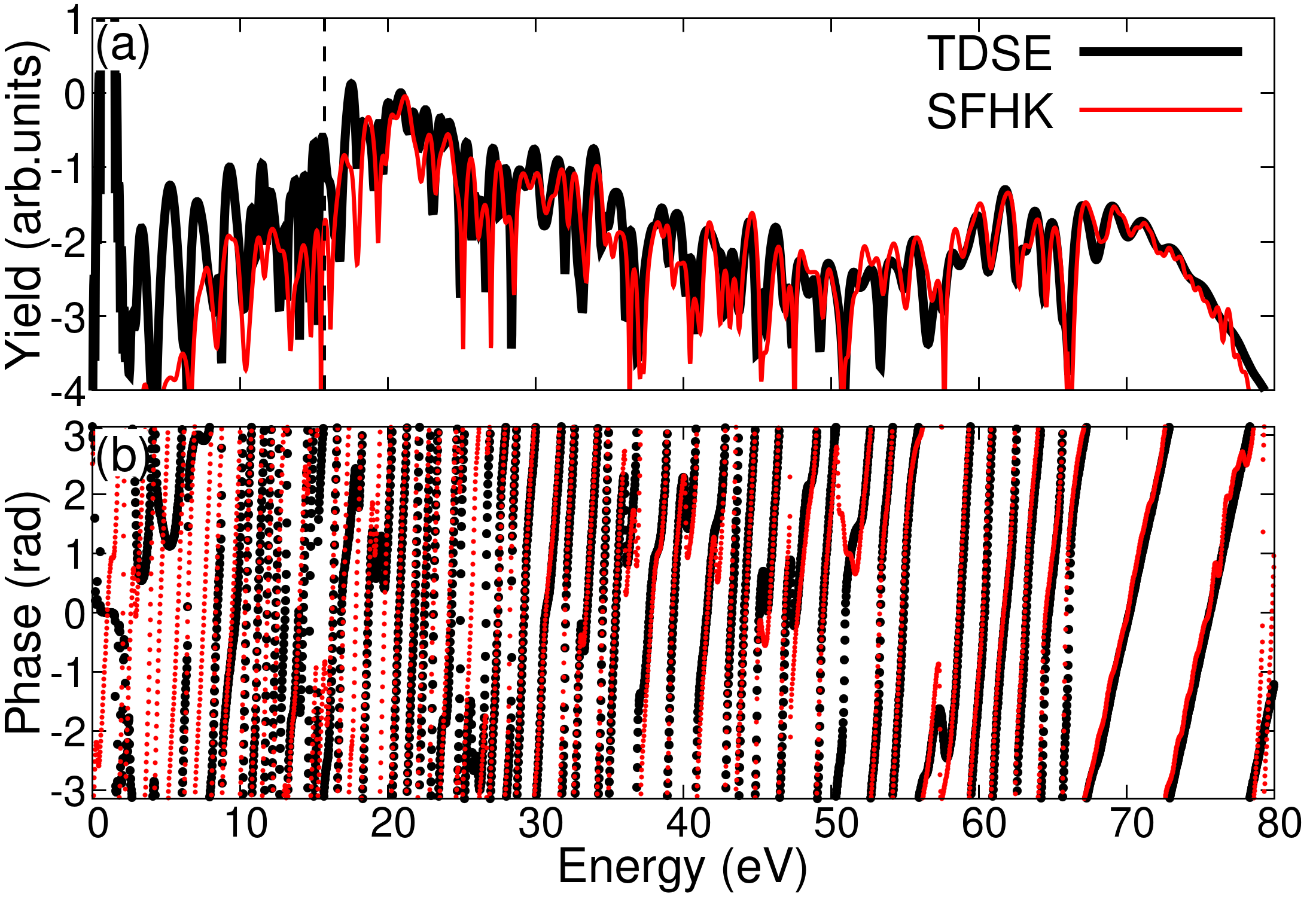}
\caption{Comparison of HHG yield (a) and phase (b), vs emitted photon energy, calculated within the SFHK and TDSE for argon. The vertical axis in (a) is on a logarithmic scale. A six-cycle laser pulse with the wavelength of 1,200-nm, intensity of $1.2 \times 10^{14}$ W/cm$^2$, and CEP $\phi=0$ was used in the calculations.}
\label{fig:Ar1200}
\end{center}
\end{figure}

In Fig.~\ref{fig:Ar1200}(a) and (b) we compare our SFHK results for the HHG yield and phase with the TDSE results. Here, we use a six-cycle pulse laser, with the wavelength of 1,200-nm, intensity of $1.2 \times 10^{14}$ W/cm$^2$, and CEP phase $\phi=0$. Note that we use the same atomic potential, proposed by Muller \cite{Muller:pra99}, in both TDSE and SFHK. The agreements between the two methods are very good, for both HHG yield and phase, in the energy range slightly above the threshold (at 15.7 eV, indicated by the dashed line in the figure). In particular, the Cooper minimum near 50 eV is nicely reproduced by the SFHK. In the case of argon, we use about 20 million trajectories, rather larger than in the hydrogen case. We speculate that this slower convergence might be due to the difficulties of the weak HHG signals in the Cooper minimum region. Note that we first calculated the ground state wavefunction of Ar($3p_0$) with the Muller potential as described in Sec.~II.2. For practical convenience, we then expressed this wavefunction through the GTOs basis sets, see Appendix B and C. 

\begin{figure}[tb!]
\begin{center}
\includegraphics[width=0.95\linewidth]{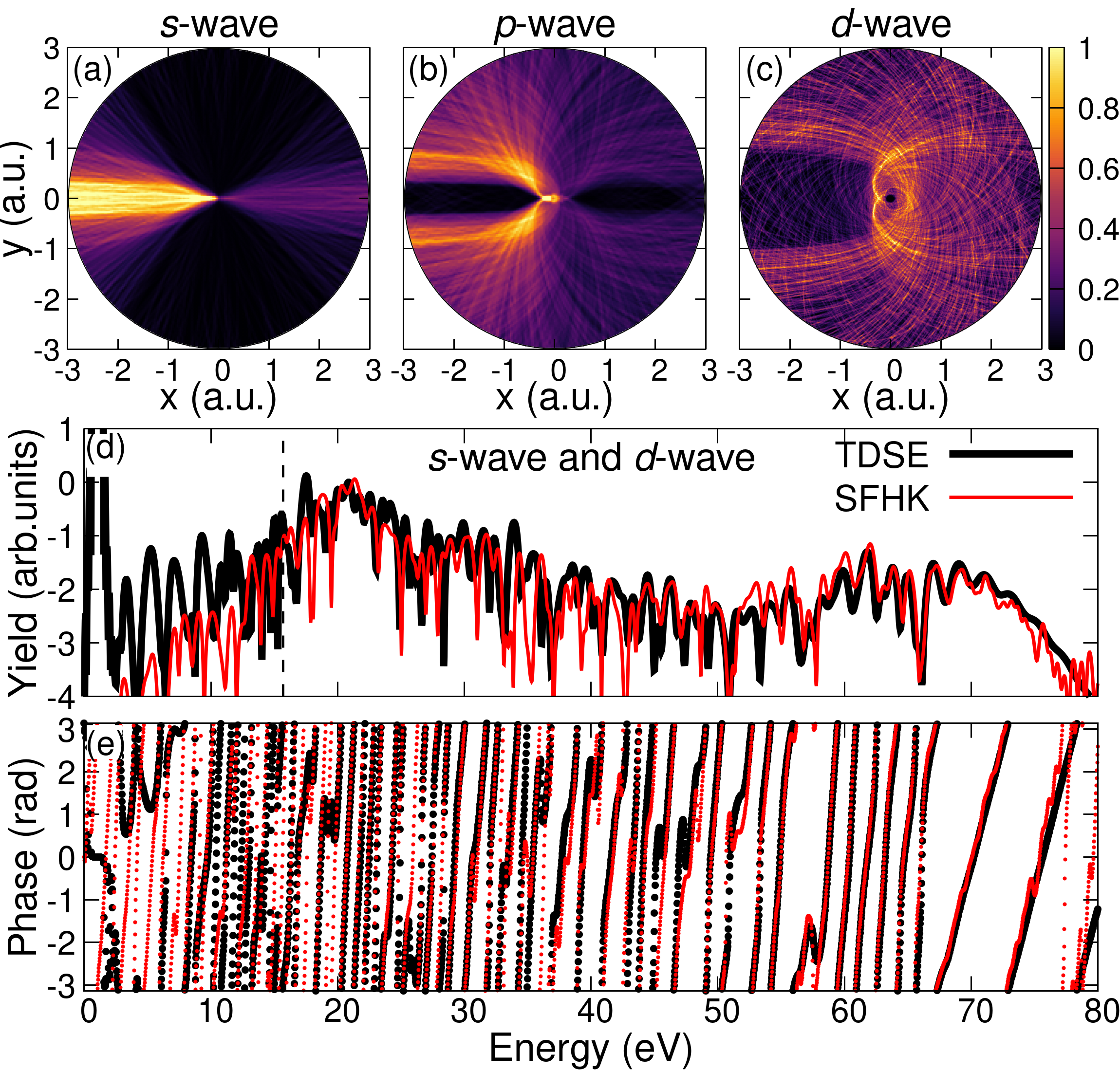}
\caption{(a), (b), and (c): Partial-wave analysis of the trajectories returning to within $r < 3$ a.u. from the core, for $s$-wave, $p$-wave, and $d$-wave, respectively, for the argon case. See text for more detail. (d) and (e): comparison of HHG yield and phase calculated with the TDSE with the SFHK, but only $s$-wave and $d$-wave trajectories that come close to within $r < 3$ a.u. from the core were included in the SFHK. The vertical axis in (d) is on a logarithmic scale. The laser parameters are the same as in Fig.~\ref{fig:Ar1200}.}
\label{fig:Arspd}
\end{center}
\end{figure}

For completeness, we also show in Fig.~\ref{fig:Arspd} the partial-wave analysis for the argon case. In this case, we expect that only $s$-wave and $d$-wave trajectories contribute to the HHG, as the initial ground state is $3p$. In fact, the SFHK calculation that includes only $s$-wave and $d$-wave trajectories that come close to within $r < 3$ a.u. from the core, agree quite well with the TDSE -- see Fig.~\ref{fig:Arspd}(d) and (e).

\subsection{On the use of tunneling theory for ionization within the SFHK} 

Having established the accuracies of the SFHK method, we now ask a question about the real needs of using the SFA for the ionization step within the SFHK. It is well-known that the SFA has a quite broad range of applicability, ranging from multiphoton regime to the tunneling regime \cite{Ivanov:jmo05}. On the other hand, in the tunneling regime, the simple quasi-static ADK theory \cite{ADK:JETP86} for atomic targets and its extension to molecular targets, such as the MO-ADK theory \cite{Tong:pra02} and the weak-field asymptotic theory(WFAT) \cite{Tolstikhin:pra11,Wahyutama:pra22}, have been shown to be rather adequate for describing ionization. Therefore, it would be of interest to see if the ADK can be used instead of the SFA to model the ionization step within the SFHK. This is important for potential applications, especially for molecules. 

\begin{figure}[tb!]
\begin{center}
\includegraphics[width=0.95\linewidth]{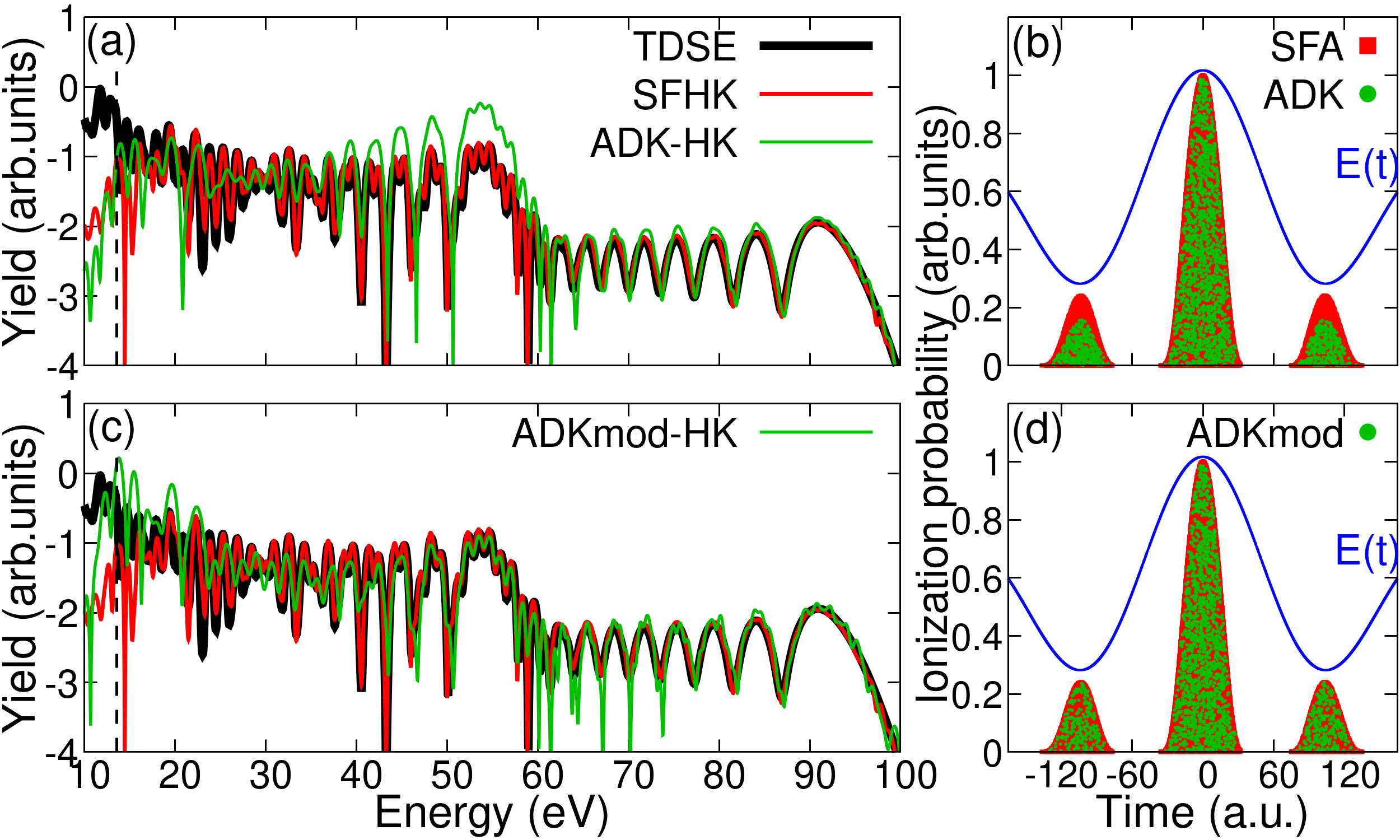}
\caption{(a): Comparison of HHG yield obtained by using the ADK-based SFHK with the TDSE result for atomic hydrogen. For reference, the HHG yield from the standard SFHK (using the SFA ionization rate) is also shown. The vertical axis is on a logarithmic scale. (b) Instantaneous ionization probability from ADK and SFA near the peak of our laser pulse. The laser electric field $E(t)$ is also shown. (c) the same as in (a) but with corrected ADK rate. (d) the same as in (b) but with the corrected ADK rate. The laser parameters are the same as in Fig.~\ref{fig:H1600}.}
\label{fig:ADK}
\end{center}
\end{figure}

To be specific, here we test on an example of atomic hydrogen for a three-cycle laser pulse with wavelength of 1,600-nm -- the same as used in Fig.~\ref{fig:H1600}. Compared to the calculation in Fig.~\ref{fig:H1600}, the only difference is that here the ADK model is used for the ionization step instead of the SFA. The HHG spectrum from this calculation, shown in Fig.~\ref{fig:ADK}(a), reveals clear discrepancies with the TDSE result. Note that we have normalized the yields to the TDSE result at the HHG cutoff energy near 90 eV. To understand the possible origin of these discrepancies, we compare in Fig.~\ref{fig:ADK}(b) the relative instantaneous ionization probabilities calculated with the ADK with that of the SFA. Clearly, the ADK underestimates ionization rate near the subcycle peaks close to $\pm 110$ a.u., as compared with the SFA rate. This is not a surprise, as the "effective" instataneous laser intensity at these two sub-peaks is almost a factor of two weaker than that at the peak of laser pulse, while the ADK is known to have a rather limited range of validity. By remormalizing the ADK rate to match the SFA rate near those two sub-peaks [see Fig.~\ref{fig:ADK}(d)], we obtained a much better agreement with the TDSE result -- see Fig.~\ref{fig:ADK}(c). This also implies that the tunneling exit used in the ADK (instead of the SFA) has minor effects on the HHG yields. Based on these results, one can generally expect that in the tunneling regime the ADK model as well as the WFAT and the MO-ADK can also be used for modeling the ionization step in the SFHK, but some proper adjustments might need to be made, especially for very short pulses.

\section{Summary and Outlook}
In conclusion, we have shown the applications of our strong-field Herman-Kluk propagator method (SFHK) to calculations of HHG spectra and induced harmonic phases. We have demonstrated very high accuracies of the SFHK for atomic hydrogen and argon in a rather broad range of laser parameters in the tunneling regime. 

We expect that extensions of our method to molecules, including polyatomic molecules, should be straightforward. In fact, the HK propagation step for molecules should not be much different from atomic targets as the most time-consuming step is essentially based on solving classical Newton equations with different initial conditions. The remaining task is to get ionization step accurately. Luckily, for atoms the SFA is quite adequate. It remains to be seen if it works well for molecules. In fact, based on the demonstrated accuracies of the HK propagator, the SFHK can be used to benchmark strong-field ionization theories, in both HHG and photoelectron momentum distribution calculations \cite{tran2024quantum}.   

Due to the semiclassical nature of our trajectory-based method, the SFHK can provide clear intuitive pictures of the HHG process. From a practical standpoint, the SFHK is also very appealing as primitive parallelization can be easily implemented for different independent trajectories.

\section*{Acknowledgments}
This work was supported by the U.S. Department of Energy (DOE), Office of Science, Basic Energy Sciences (BES) under Award Number DE-SC0023192. We thank Richard Jones and the Storrs HPC facilities for valuable computational supports and resources. This research was done using services provided by the OSG Consortium \cite{osg07new,osg09new,osg06new,osg05new}.

\appendix
\onecolumngrid
\setcounter{equation}{0}
\renewcommand{\theequation}{\Alph{section}.\arabic{equation}}

\section{Continuum wavefunction constructed within the SFHK}

The outgoing wavefunction $\Delta\psi_c ({\bf{p}},t)$ can be written as a propagation of the continuum wavefunction $\Delta\psi_c({\bf{k}},t_0)$, which is evaluated immediately after tunneling at time $t_0$, by exact quantum propagator $K({\bf{p}},t;{\bf{k}},t_0)$,
\begin{equation}
\Delta\psi_c({\bf{p}},t) = \int d{\bf{k}} K({\bf{p}},t;{\bf{k}},t_0) \Delta\psi_c({\bf{k}},t_0).
\label{ExactEq}
\end{equation}

Here, $\Delta\psi_c({\bf{k}},t_0)$ is the probability amplitude of an active electron tunneled from the ground state $\left| \psi_0 \right\rangle$ to the continuum state $\left| {\bf{k}} \right\rangle$ during a small time window $\Delta t$ just preceding time $t_0$. It can be described by SFA theory \cite{Ivanov:jmo05,Milosevic:jpb06,Lai2015,Brennecke2019,Brennecke2020}, leading to
\begin{equation}
\begin{split}
\Delta\psi_c({\bf{k}},t_0) & \propto -i  \sqrt{\dfrac{2\pi i}{\partial^2 S_\downarrow^0({\bf{k}},t_s)/\partial t_s^2}} \left\langle\ {\bf{k}} + {\bf{A}}(t_s) -  {\bf{A}}(t_0) \right| {\bf{r}} \cdot {\bf{E}}(t_s) \left| \psi_0 \right\rangle e^{iS_\downarrow^0({\bf{k}},t_s)} \\ 
& = -i \sqrt{\dfrac{2\pi i}{{\bf{E}}(t_s)\cdot [{\bf{k}} + {\bf{A}}(t_s) -  {\bf{A}}(t_0)] }} \left\langle\ {\bf{k}} + {\bf{A}}(t_s) -  {\bf{A}}(t_0) \right| {\bf{r}} \cdot {\bf{E}}(t_s) \left| \psi_0 \right\rangle e^{iS_\downarrow^0({\bf{k}},t_s)},
\end{split}
\label{SFAwf}
\end{equation}
where ${\bf{A}}(t)$ is the vector potential and ${\bf{E}}(t)=-\partial_t {\bf{A}}(t)$ is the electric field of a laser pulse defined by Eq.(\ref{EfieldTDSE}). $t_s = t_0 + it_t$ is the complex-valued solution of the saddle-point equation
\begin{equation}
\frac{[{\bf{k}} + {\bf{A}}(t_s) -  {\bf{A}}(t_0)]^2}{2}+I_p = 0.
\label{SAPeq}
\end{equation}
Note that $\left| {\bf{k}} \right\rangle$ forms a plane wave basis and the wavefunction $\Delta \psi_c({\bf{k}},t_0)$ has been approximated by Volkov states only during the tunneling through the potential barier. The phase in (\ref{SFAwf}) is the action during tunneling and it is given by $S_\downarrow^0({\bf{k}},t_s) = I_pt_s-\int_{t_s}^{t_0}\frac{[{\bf{k}} + {\bf{A}}(t'') -  {\bf{A}}(t_0)]^2}{2} dt''$.

In the SFHK method, the exact propagator in Eq.(\ref{ExactEq}) is approximated by the Herman-Kluk propagator \cite{Herman1984, Kluk1986, Kay1994_1, Kay1994_2, Kay1994_3}, which is
\begin{equation}
K^{HK}({\bf{p}},t;{\bf{k}},t_0)=\dfrac{1}{(2\pi)^3}\iint d{\bf{p}}_{t_0} d{\bf{q}}_{t_0} C({\bf{p}}_{t_0},{\bf{q}}_{t_0},t)e^{iS_\to({\bf{p}}_{t_0},{\bf{q}}_{t_0},t)}\braket{{\bf{p}}|{\bf{p}}_t,{\bf{q}}_t,\gamma}\braket{{\bf{p}}_{t_0},{\bf{q}}_{t_0},\gamma|{\bf{k}}},
\label{HKpropa}
\end{equation}
where $\braket{{\bf{p}}|{\bf{p}}_t,{\bf{q}}_t,\gamma} = (\frac{1}{2\pi \gamma})^{3/4} e^{-\frac{({\bf{p}} - {\bf{p}}_t)^2}{4\gamma}-i{\bf{p}} \cdot {\bf{q}}_t}$ is the coherent state wavefunction with the width $\gamma$, the average momentum ${\bf{p}}_t$, and the average position ${\bf{q}}_t$, associated with the phase space $({\bf{p}}_t,{\bf{q}}_t)$ of a trajectory at time $t$. The propagation of an individual trajectory from initial conditions $({\bf{p}}_{t_0},{\bf{q}}_{t_0})$ at time $t_0$ is governed by the Hamilton's equations of motion
\begin{equation}
{\bf{\dot q}}_t = \dfrac{\partial H}{\partial {\bf{p}}_t}, \ {\bf{\dot p}}_t = -\dfrac{\partial H}{\partial {\bf{q}}_t}.
\end{equation} 
Here, the Hamiltonian is given by $H = \frac{{\bf{p}}_{t}^2}{2}+V_a({\bf{q}}_{t})+{\bf{q}}_{t} \cdot {\bf{E}}(t)$, where $V_a$ is the atomic potential between a single active electron and the parent ion, and the action reads $S_\to({\bf{p}}_{t_0},{\bf{q}}_{t_0},t) = \int_{t_0}^t ({\bf{p}}_{t'} \cdot {\bf{\dot q}}_{t'} - H) dt'$. The complex-valued prefactor proposed by Herman-Kluk \cite{Herman1984, Kluk1986} can be written as
\begin{equation}
C({\bf{p}}_{t_0},{\bf{q}}_{t_0},t) = \sqrt{\det \left[\dfrac{1}{2} \left(\dfrac{\partial {\bf{p}}_t}{\partial {\bf{p}}_{t_0}} + \dfrac{\partial {\bf{q}}_t}{\partial {\bf{q}}_{t_0}} - 2i\gamma \dfrac{\partial {\bf{q}}_t}{\partial {\bf{p}}_{t_0}} - \dfrac{1}{2i\gamma}\dfrac{\partial {\bf{p}}_t}{\partial {\bf{q}}_{t_0}} \right) \right]}.
\label{C_HK}
\end{equation}
The stability matrix elements appearing in Eq.(\ref{C_HK})   fullfill equations \cite{Kay1994_3}
\begin{equation}
\begin{split}
& \dfrac{d}{dt}\left(\dfrac{\partial p_{ti}}{\partial z_j} \right) = -\sum_{k=1}^{3} \left( \dfrac{\partial^2 H}{\partial p_{tk} \partial q_{ti}} \dfrac{\partial p_{tk}}{\partial z_j} + \dfrac{\partial^2 H}{\partial q_{tk} \partial q_{ti}} \dfrac{\partial q_{tk}}{\partial z_j} \right), \\
& \dfrac{d}{dt}\left(\dfrac{\partial q_{ti}}{\partial z_j} \right) = \sum_{k=1}^{3} \left( \dfrac{\partial^2 H}{\partial p_{tk} \partial p_{ti}} \dfrac{\partial p_{tk}}{\partial z_j} + \dfrac{\partial^2 H}{\partial q_{tk} \partial p_{ti}} \dfrac{\partial q_{tk}}{\partial z_j} \right), \\
& z = p_{t_0} \ {\rm{or}} \ q_{t_0}.
\end{split}
\end{equation}

By substitutiing (\ref{SFAwf}) and (\ref{HKpropa}) into Eq.~(\ref{ExactEq}) we get
\begin{equation}
\begin{split}
\Delta \psi_c ({\bf{p}},t) & \propto -i \int d{\bf{k}} \dfrac{1}{(2\pi)^3}\iint d{\bf{p}}_{t_0} d{\bf{q}}_{t_0} C({\bf{p}}_{t_0},{\bf{q}}_{t_0},t)e^{iS_\to({\bf{p}}_{t_0},{\bf{q}}_{t_0},t)}\braket{{\bf{p}}|{\bf{p}}_t,{\bf{q}}_t,\gamma}\braket{{\bf{p}}_{t_0},{\bf{q}}_{t_0},\gamma|{\bf{k}}} \\
& \ \ \ \ \ \ \ \ \ \ \ \ \ \ \times \sqrt{\dfrac{2\pi i}{{\bf{E}}(t_s)\cdot [{\bf{k}} + {\bf{A}}(t_s) -  {\bf{A}}(t_0)] }} \left\langle\ {\bf{k}} + {\bf{A}}(t_s) -  {\bf{A}}(t_0) \right| {\bf{r}} \cdot {\bf{E}}(t_s) \left| \psi_0 \right\rangle e^{iS_\downarrow^0({\bf{k}},t_s)} \\
& \propto \iint d{\bf{p}}_{t_0} d{\bf{q}}_{t_0} C({\bf{p}}_{t_0},{\bf{q}}_{t_0},t)e^{iS_\to({\bf{p}}_{t_0},{\bf{q}}_{t_0},t)}\braket{{\bf{p}}|{\bf{p}}_t,{\bf{q}}_t,\gamma} \\
& \ \ \ \ \ \   \times \int d{\bf{k}} \braket{{\bf{p}}_{t_0},{\bf{q}}_{t_0},\gamma|{\bf{k}}} \left\langle\ {\bf{k}} + {\bf{A}}(t_s) -  {\bf{A}}(t_0) \right| {\bf{r}} \cdot {\bf{E}}(t_s) \left| \psi_0 \right\rangle \dfrac{e^{iS_\downarrow^0({\bf{k}},t_s)}}{\sqrt{{\bf{E}}(t_s)\cdot [{\bf{k}} + {\bf{A}}(t_s) -  {\bf{A}}(t_0)] }}.
\end{split}
\label{SFHK1}
\end{equation}

Integration over $\bf{k}$ in Eq.~({\ref{SFHK1}}) can be done by realizing that $\braket{{\bf{k}}|{\bf{p}},{\bf{q}},\gamma} = (\frac{1}{2\pi \gamma})^{3/4}e^{-\frac{({\bf{k}}-{\bf{p}})^2}{4\gamma} -i{\bf{k}} \cdot {\bf{q}}}$ and using an approximation of $\lim\limits_{\gamma\to0} (\frac{1}{4\pi \gamma})^{3/2}e^{-\frac{({\bf{k}}-{\bf{p}})^2}{4\gamma}} = \delta({\bf{k}}-{\bf{p}})$. Then for small $\gamma$, one gets
\begin{equation}
\begin{split}
I_0 & = \int d{\bf{k}} \braket{{\bf{p}}_{t_0},{\bf{q}}_{t_0},\gamma|{\bf{k}}} \left\langle\ {\bf{k}} + {\bf{A}}(t_s) -  {\bf{A}}(t_0) \right| {\bf{r}} \cdot {\bf{E}}(t_s) \left| \psi_0 \right\rangle \dfrac{e^{iS_\downarrow^0({\bf{k}},t_s)}}{\sqrt{{\bf{E}}(t_s)\cdot [{\bf{k}} + {\bf{A}}(t_s) -  {\bf{A}}(t_0)] }} \\
& \approx \int d{\bf{k}} \delta({\bf{k}} - {\bf{p}}_{t_0}) e^{i{\bf{k}} \cdot {\bf{q}}_{t_0}}  \left\langle\ {\bf{k}} + {\bf{A}}(t_s) -  {\bf{A}}(t_0) \right| {\bf{r}} \cdot {\bf{E}}(t_s) \left| \psi_0 \right\rangle \dfrac{e^{iS_\downarrow^0({\bf{k}},t_s)}}{\sqrt{{\bf{E}}(t_s)\cdot [{\bf{k}} + {\bf{A}}(t_s) -  {\bf{A}}(t_0)] }} \\
& = e^{i{\bf{p}}_{t_0} \cdot {\bf{q}}_{t_0}}  \left\langle\ {\bf{p}}_{t_0} + {\bf{A}}(t_s) -  {\bf{A}}(t_0) \right| {\bf{r}} \cdot {\bf{E}}(t_s) \left| \psi_0 \right\rangle \dfrac{e^{iS_\downarrow^0({\bf{p}}_{t_0},t_s)}}{\sqrt{{\bf{E}}(t_s)\cdot [{\bf{p}}_{t_0} + {\bf{A}}(t_s) -  {\bf{A}}(t_0)] }}.
\end{split}
\label{Int}
\end{equation}
The integral (\ref{Int}) implies that the initial momentum immediately after tunneling is given by ${\bf{p}}_{t_0} = \bf{k}$, where $\bf{k}$ is the momentum at the tunnel exit -- see Eq.~(\ref{SAPeq}). The initial position which is the ``tunnel exit" point can be written as ${\bf{q}}_{t_0} = {\rm Re}[\int_{t_s}^{t_0}{\bf{A}}(\tau) d\tau] $ as shown in Refs. \cite{Yan:prl2010,Lai2015,Brennecke2020}. Therefore, $({\bf{p}}_{t_0},{\bf{q}}_{t_0})$ are chosen by a Monte Carlo algorithm when ${\bf{k}}$ is generated randomly. A ${\bf{k}}$ that fullfills the saddle-point equation (\ref{SAPeq}) provides a weight $\propto |I_0 |$ to implement the importance sampling in the Monte Carlo integration. We note that $I_0$ is the continuum wavefunction immediately after tunneling in the coherent states representation. In other words, $|I_0 |^2$ is proportional to Husimi distribution at the tunnel exit time.

By substituting (\ref{Int}) into Eq.~(\ref{SFHK1}) and adding the contributions from all possible ionization times, we finally get \cite{tran2024quantum,wcl3-x52t,mcmanus2025delay}
\begin{equation}
\begin{split}
\psi_c ({\bf{p}},t) &  \propto \sum_s \iint d{\bf{p}}_{t_0} d{\bf{q}}_{t_0} \braket{{\bf{p}}|{\bf{p}}_t,{\bf{q}}_t,\gamma} C({\bf{p}}_{t_0},{\bf{q}}_{t_0},t) e^{i(S_\to + S_\downarrow^0 + {\bf{p}}_{t_0} \cdot {\bf{q}}_{t_0})} \\
&\ \ \ \ \ \  \ \ \ \ \ \ \ \  \times \dfrac{\left\langle\ {\bf{p}}_{t_0} + {\bf{A}}(t_s) -  {\bf{A}}(t_0) \right| {\bf{r}} \cdot {\bf{E}}(t_s) \left| \psi_0 \right\rangle}{\sqrt{{\bf{E}}(t_s)\cdot [{\bf{p}}_{t_0} + {\bf{A}}(t_s) -  {\bf{A}}(t_0)] }}.
\end{split}
\label{SFHKwf}
\end{equation}

\section{On the calculations of tunneling matrix elements}
\subsection{Tunneling matrix elements and the expansion in the Gaussian-type orbitals (GTOs)}

In general, the initial ground states of atoms or molecules can be expanded into the Gaussian-type orbitals (GTOs) basis sets \cite{Hehre:jcp69,Ditchfield:jcp71}. In this paper, we fit the ground state wavefunction obtained by the time-independent Schr\"odinger equation to a GTOs basis set. For example, in this work, we choose 6-311G and cc-pV5Z as the GTOs basis sets for hydrogen $1s$ and argon $3p$, respectively. The exponents of GTOs are remained unchanged. The coefficients of GTOs are obtained by linear regression.  For simplicity, we illustrate the case in which only the $s$-wave is included in the GTOs basis set, $\psi_0({\bf{r}}) = \sum\limits_j c_je^{-\alpha_j{\bf{r}}^2}$ (e.g., 6-311G in hydrogen), and the laser pulse is polarized linearly along the z-axis. Then the tunneling matrix elements used in the SFHK can be written as
\begin{equation}
\begin{split}
\left\langle\ {\bf{k}} \right| {\bf{r}} \cdot {\bf{E}}(t) \left| \psi_0 \right\rangle & = E_z(t)\left\langle\ {\bf{k}} \right| z \left| \psi_0 \right\rangle \\
& = E_z(t) \int d{\bf{r}} \ \braket{{\bf{k}}|{\bf{r}}} z \braket{{\bf{r}}|\psi_0} \\
& = E_z(t) \int d{\bf{r}} e^{-i {\bf{k}} \cdot {\bf{r}}} z \sum\limits_j c_je^{-\alpha_j{\bf{r}}^2} \\
& \propto E_z(t) \sum\limits_j c_j \int e^{-\alpha_jx^2-ik_xx}dx \int e^{-\alpha_jy^2-ik_yy} dy \int z e^{-\alpha_jz^2-ik_zz}dz \\
& \propto E_z(t) k_z \sum\limits_j \frac{c_j}{\sqrt{\alpha_j^5}}e^{\frac{-{\bf{k}}^2}{4\alpha_j}}.
\end{split}
\end{equation}
By changing variables, one gets
\begin{equation}
\left\langle\ {\bf{p}}_{t_0} + {\bf{A}}(t_s) -  {\bf{A}}(t_0) \right| {\bf{r}} \cdot {\bf{E}}(t) \left| \psi_0 \right\rangle \propto E_z(t) [p_{z_{t_0}} + A_z(t_s) - A_z(t_0)] \sum\limits_j \frac{c_j}{\sqrt{\alpha_j^5}}e^{\frac{-[{\bf{p}}_{t_0} + {\bf{A}}(t_s) -  {\bf{A}}(t_0)]^2}{4\alpha_j}}.
\end{equation}
For $p$-wave GTOs, $\psi_0({\bf{r}}) = z\sum\limits_j c_je^{-\alpha_j{\bf{r}}^2}$, the tunneling matrix elements read
\begin{equation}
\left\langle\ {\bf{k}} \right| {\bf{r}} \cdot {\bf{E}}(t) \left| \psi_0 \right\rangle \propto E_z(t) \sum\limits_j \frac{c_j}{\sqrt{\alpha_j^7}}e^{\frac{-{\bf{k}}^2}{4\alpha_j}} (-k_z^2+2\alpha_j).
\end{equation}

\subsection{The modified saddle-point method}

For the Coulomb potential, the tunneling matrix elements in Eq.(\ref{SFHKwf}) become singular \cite{Milosevic:jpb06}. In that case, one can apply the modified saddle-point method as in Ref.~\cite{Milosevic:jpb06}, which provides the continuum wavefunction after tunneling as
\begin{equation}
\psi({\bf{k}},t_0) \propto \sum_s \dfrac{e^{iS_\downarrow^0({\bf{k}},t_s)}}{{\bf{E}}(t_s)\cdot [{\bf{k}} + {\bf{A}}(t_s) -  {\bf{A}}(t_0)] }.
\end{equation}

Then the SFHK wavefunction (\ref{SFHKwf}) can be simplified as \cite{tran2024quantum,wcl3-x52t,mcmanus2025delay}
\begin{equation}
\psi ({\bf{p}},t) \propto \sum_s \iint d{\bf{p}}_{t_0} d{\bf{q}}_{t_0} \braket{{\bf{p}}|{\bf{p}}_t,{\bf{q}}_t,\gamma} C({\bf{p}}_{t_0},{\bf{q}}_{t_0},t) \dfrac{e^{i(S_\to + S_\downarrow^0 + {\bf{p}}_{t_0} \cdot {\bf{q}}_{t_0})}}{{\bf{E}}(t_s)\cdot [{\bf{p}}_{t_0} + {\bf{A}}(t_s) -  {\bf{A}}(t_0)] }.
\label{SFHKmodSPA}
\end{equation}

\section{Calculations of dipole and dipole velocity in the GTO basis}

\subsection{Dipole}

In SFHK, the dipole moment along the laser polarization axis can be expressed as
\begin{equation}
\begin{split}
D_z(t) & = - \left\langle\ \Psi(t) \right| z \left| \Psi(t) \right\rangle \\
& = - \left\langle\ \psi_0e^{iI_pt} + \psi_c(t) \right| z \left| \psi_0e^{iI_pt} + \psi_c(t) \right\rangle \\
& \approx - \left\langle\ \psi_0e^{iI_pt} \right| z \left| \psi_c(t) \right\rangle + c.c. \\
& \propto - \sum_s \iint d{\bf{p}}_{t_0} d{\bf{q}}_{t_0} e^{iI_pt}\left\langle\ \psi_0 \right| z \left| {\bf{p}}_t,{\bf{q}}_t,\gamma \right\rangle C({\bf{p}}_{t_0},{\bf{q}}_{t_0},t) e^{i(S_\to + S_\downarrow^0 + {\bf{p}}_{t_0} \cdot {\bf{q}}_{t_0})} \\
&\ \ \ \ \ \  \ \ \ \ \ \ \ \  \times \dfrac{\left\langle\ {\bf{p}}_{t_0} + {\bf{A}}(t_s) -  {\bf{A}}(t_0) \right| {\bf{r}} \cdot {\bf{E}}(t_s) \left| \psi_0 \right\rangle}{\sqrt{{\bf{E}}(t_s)\cdot [{\bf{p}}_{t_0} + {\bf{A}}(t_s) -  {\bf{A}}(t_0)] }} + c.c.
\end{split}
\label{D_t}
\end{equation} 

For $s$-wave, by using the same approach as described above for the GTOs expansion, the dipole moment matrix elements in Eq.(\ref{D_t}) can be written as
\begin{equation}
\begin{split}
\left\langle\ \psi_0 \right| z \left| {\bf{p}}_t,{\bf{q}}_t,\gamma \right\rangle & = \int d{\bf{r}} \braket{\psi_0|{\bf{r}}} z\braket{{\bf{r}}|{\bf{p}}_t,{\bf{q}}_t,\gamma} \\ 
& \propto \int d{\bf{r}} \sum\limits_j c_je^{-\alpha_j{\bf{r}}^2} z e^{-\gamma({\bf{r}}-{\bf{q}}_t)^2 +i {\bf{p}}_t \cdot ({\bf{r}}-{\bf{q}}_t)} \\
& \propto \sum\limits_j c_j e^{-\gamma{\bf{q}}_t^2 - i{\bf{p}}_t \cdot {\bf{q}}_t} \int e^{-(\gamma+\alpha_j)x^2+(2\gamma q_x+ip_x)x} dx \\ 
& \ \ \ \ \ \ \ \ \ \ \ \ \ \ \ \ \ \ \ \ \ \ \ \ \times \int e^{-(\gamma+\alpha_j)y^2+(2\gamma q_y+ip_y)y} dy \\
& \ \ \ \ \ \ \ \ \ \ \ \ \ \ \ \ \ \ \ \ \ \ \ \ \times\int z e^{-(\gamma+\alpha_j)z^2+(2\gamma q_z+ip_z)z} dz \\
& \propto (2\gamma q_z+ip_z) \sum\limits_j \dfrac{c_j}{\sqrt{(\gamma+\alpha_j)^5}}e^{\frac{-{\bf{p}}_t^2 - 4\alpha_j(\gamma{\bf{q}}_t^2 + i{\bf{p}}_t \cdot {\bf{q}}_t)}{4(\gamma+\alpha_j)}}.
\end{split}
\end{equation}

For $p$-wave GTOs, the dipole moment becomes 
\begin{equation}
\left\langle\ \psi_0 \right| z \left| {\bf{p}}_t,{\bf{q}}_t,\gamma \right\rangle \propto \sum\limits_j \dfrac{c_j}{\sqrt{(\gamma+\alpha_j)^7}}e^{\frac{-{\bf{p}}_t^2 - 4\alpha_j(\gamma{\bf{q}}_t^2 + i{\bf{p}}_t \cdot {\bf{q}}_t)}{4(\gamma+\alpha_j)}} [(2\gamma q_z+ip_z)^2+2(\gamma+\alpha_j)].
\end{equation}

\subsection{Dipole velocity}
Similarly to the dipole case, the dipole velocity constructed by SFHK can be written as
\begin{equation}
\begin{split}
v_z(t) & = - \left\langle\ \Psi(t) \right| k_z \left| \Psi(t) \right\rangle \\
& \propto - \sum_s \iint d{\bf{p}}_{t_0} d{\bf{q}}_{t_0} e^{iI_pt}\left\langle\ \psi_0 \right| k_z \left| {\bf{p}}_t,{\bf{q}}_t,\gamma \right\rangle C({\bf{p}}_{t_0},{\bf{q}}_{t_0},t) e^{i(S_\to + S_\downarrow^0 + {\bf{p}}_{t_0} \cdot {\bf{q}}_{t_0})} \\
&\ \ \ \ \ \  \ \ \ \ \ \ \ \  \times \dfrac{\left\langle\ {\bf{p}}_{t_0} + {\bf{A}}(t_s) -  {\bf{A}}(t_0) \right| {\bf{r}} \cdot {\bf{E}}(t_s) \left| \psi_0 \right\rangle}{\sqrt{{\bf{E}}(t_s)\cdot [{\bf{p}}_{t_0} + {\bf{A}}(t_s) -  {\bf{A}}(t_0)] }} + c.c.
\end{split}
\label{V_t}
\end{equation}

For $s$-wave Gaussian basis set, the dipole velocity matrix elements is
\begin{equation}
\begin{split}
\left\langle\ \psi_0 \right| k_z \left| {\bf{p}}_t,{\bf{q}}_t,\gamma \right\rangle & = \int d{\bf{k}} \braket{\psi_0|{\bf{k}}} k_z\braket{{\bf{k}}|{\bf{p}}_t,{\bf{q}}_t,\gamma} \\ 
& \propto \int d{\bf{k}} \sum\limits_j \dfrac{c_j}{\sqrt{\alpha_j^3}} e^{- \frac{{\bf{k}}^2}{4\alpha_j}} k_z e^{- \frac{({\bf{k}}-{\bf{p}}_t)^2}{4\gamma} - i {\bf{k}} \cdot {\bf{q}}_t} \\
& \propto \sum\limits_j \dfrac{c_j}{\sqrt{\alpha_j^3}} e^{-\frac{{\bf{p}}_t^2}{4\gamma}} \int e^{ \frac{-(\gamma+\alpha_j)k_x^2+(2\alpha_j p_x-4i\gamma \alpha_jq_x)k
_x}{4\gamma\alpha_j} } dk_x \\ 
& \ \ \ \ \ \ \ \ \ \ \ \ \ \ \ \ \ \ \ \times \int e^{ \frac{-(\gamma+\alpha_j)k_y^2+(2\alpha_j p_y-4i\gamma \alpha_jq_y)k
_y}{4\gamma\alpha_j} } dk_y \\
& \ \ \ \ \ \ \ \ \ \ \ \ \ \ \ \ \ \ \ \times \int k_z e^{ \frac{-(\gamma+\alpha_j)k_z^2+(2\alpha_j p_z-4i\gamma \alpha_jq_z)k
_z}{4\gamma\alpha_j} } dk_z \\
& \propto (p_z-2i\gamma q_z) \sum\limits_j \dfrac{c_j\alpha_j}{\sqrt{(\gamma+\alpha_j)^5}}e^{\frac{-{\bf{p}}_t^2 - 4\alpha_j(\gamma{\bf{q}}_t^2 + i{\bf{p}}_t \cdot {\bf{q}}_t)}{4(\gamma+\alpha_j)}}.
\end{split}
\end{equation}

For $p$-wave basis set, one gets
\begin{equation}
\left\langle\ \psi_0 \right| k_z \left| {\bf{p}}_t,{\bf{q}}_t,\gamma \right\rangle \propto \sum\limits_j \dfrac{c_j\alpha_j}{\sqrt{(\gamma+\alpha_j)^7}}e^{\frac{-{\bf{p}}_t^2 - 4\alpha_j(\gamma{\bf{q}}_t^2 + i{\bf{p}}_t \cdot {\bf{q}}_t)}{4(\gamma+\alpha_j)}} \left[ \dfrac{(p_z-2i\gamma q_z)^2}{2\gamma} + \dfrac{\gamma+\alpha_j}{\alpha_j} \right].
\end{equation}

\section{Calculations of HHG spectra}

Once the time-dependent induced dipole or dipole velocity is obtained, the HHG spectrum can be calculated using the Fourier transforms

\begin{equation}
\begin{split}
& D_z(\omega) = \int D_z(t) W_H(t) e^{i\omega t} dt, \\ 
& v_z(\omega) = \int v_z(t) W_H(t) e^{i\omega t} dt, 
\end{split}
\end{equation}
where $W_H(t) = 0.5\left[1+\cos \left(\frac{\pi t}{\tau_H}\right)\right]$ is the Hanning window function used to drop the dipole smoothly to zero at the end of laser pulse \cite{Camp:jpb18}. Therefore, the width, $\tau_H$, of the window function is chosen such that $W_H(t) = 0$ when the laser pulse ends. We confirm that the yield of HHG spectra shown in this work are proportional to $|D_z(\omega)|^2 \omega^4 \approx |v_z(\omega)|^2 \omega^2$ as derived in Ref.~\cite{BandraukPRA2009}. We note that the same Hanning window is applied to both TDSE and SFHK dipole or dipole velocity.

\bibliography{MyBib,Phi_Bib} 

@Article{Corkum:prl93,
  title = {Plasma perspective on strong field multiphoton ionization},
  author = {Corkum, P. B.},
  journal = {Phys. Rev. Lett.},
  volume = {71},
  number = {13},
  pages = {1994--1997},
  numpages = {3},
  year = {1993},
  month = {Sep},
  doi = {10.1103/PhysRevLett.71.1994},
  publisher = {American Physical Society}
}

@article{Lewenstein:pra94,
  title = {Theory of high-harmonic generation by low-frequency laser fields},
  author = {Lewenstein, M.  and Balcou, Ph.  and Ivanov, M. Yu. and L 'Huillier,
  Anne  and Corkum, P. B.},
  journal = {Phys. Rev. A},
  volume = {49},
  number = {3},
  pages = {2117--2132},
  numpages = {15},
  year = {1994},
  month = {Mar},
  doi = {10.1103/PhysRevA.49.2117},
  publisher = {American Physical Society}
}

@book{Lin:book2018, 
    place={Cambridge}, 
    title={Attosecond and Strong-Field Physics: Principles and Applications}, 
    DOI={10.1017/9781108181839}, publisher={Cambridge University Press}, 
    author={Lin, C. D. and Le, Anh-Thu and Jin, Cheng and Wei, Hui}, 
    year={2018}
}

@article{Milosevic:jpb06,
  author={D B Milosevic and G G Paulus and D Bauer and W Becker},
  title={Above-threshold ionization by few-cycle pulses},
  journal={J. Phys. B},
  volume={39},
  number={14},
  pages={R203-R262},
  url={http://stacks.iop.org/0953-4075/39/R203},
  year={2006}
}

@article{Faria:PhysRep20,
doi = {10.1088/1361-6633/ab5c91},
url = {https://dx.doi.org/10.1088/1361-6633/ab5c91},
year = {2020},
month = {jan},
publisher = {IOP Publishing},
volume = {83},
number = {3},
pages = {034401},
author = {C Figueira de Morisson Faria and A S Maxwell},
title = {It is all about phases: ultrafast holographic photoelectron imaging},
journal = {Rep. Prog. Phys.}
}

@article{Yan:prl10,
  title = {Low-Energy Structures in Strong Field Ionization Revealed by Quantum Orbits},
  author = {Yan, Tian-Min and Popruzhenko, S. V. and Vrakking, M. J. J. and Bauer, D.},
  journal = {Phys. Rev. Lett.},
  volume = {105},
  issue = {25},
  pages = {253002},
  numpages = {4},
  year = {2010},
  month = {Dec},
  publisher = {American Physical Society},
  doi = {10.1103/PhysRevLett.105.253002},
  url = {https://link.aps.org/doi/10.1103/PhysRevLett.105.253002}
}

@article{Shvetsov:pra16,
  title = {Semiclassical two-step model for strong-field ionization},
  author = {Shvetsov-Shilovski, N. I. and Lein, M. and Madsen, L. B. and R\"as\"anen, E. and Lemell, C. and Burgd\"orfer, J. and Arb\'o, D. G. and T\"okesi, K.},
  journal = {Phys. Rev. A},
  volume = {94},
  issue = {1},
  pages = {013415},
  numpages = {12},
  year = {2016},
  month = {Jul},
  publisher = {American Physical Society},
  doi = {10.1103/PhysRevA.94.013415},
  url = {https://link.aps.org/doi/10.1103/PhysRevA.94.013415}
}

@article{Brennecke2019,
  title = {Strong-field photoelectron holography beyond the electric dipole approximation: A semiclassical analysis},
  author = {Brennecke, Simon and Lein, Manfred},
  journal = {Phys. Rev. A},
  volume = {100},
  issue = {2},
  pages = {023413},
  numpages = {13},
  year = {2019},
  month = {Aug},
  publisher = {American Physical Society},
  doi = {10.1103/PhysRevA.100.023413},
  url = {https://link.aps.org/doi/10.1103/PhysRevA.100.023413}
}

@article{Brennecke2020,
  title = {Gouy's Phase Anomaly in Electron Waves Produced by Strong-Field Ionization},
  author = {Brennecke, Simon and Eicke, Nicolas and Lein, Manfred},
  journal = {Phys. Rev. Lett.},
  volume = {124},
  issue = {15},
  pages = {153202},
  numpages = {7},
  year = {2020},
  month = {Apr},
  publisher = {American Physical Society},
  doi = {10.1103/PhysRevLett.124.153202},
  url = {https://link.aps.org/doi/10.1103/PhysRevLett.124.153202}
}

@article{Herman1984,
title = {A semiclasical justification for the use of non-spreading wavepackets in dynamics calculations},
journal = {Chem. Phys.},
volume = {91},
number = {1},
pages = {27-34},
year = {1984},
doi = {https://doi.org/10.1016/0301-0104(84)80039-7},
url = {https://www.sciencedirect.com/science/article/pii/0301010484800397},
author = {Michael F. Herman and Edward Kluk}
}

@article{Kluk1986,
    author = {Kluk, Edward and Herman, Michael F. and Davis, Heidi L.},
    title = "{Comparison of the propagation of semiclassical frozen Gaussian wave functions with quantum propagation for a highly excited anharmonic oscillator}",
    journal = {J. Chem. Phys.},
    volume = {84},
    number = {1},
    pages = {326-334},
    year = {1986},
    month = {01},
    doi = {10.1063/1.450142},
    url = {https://doi.org/10.1063/1.450142}
}

@article{Le:pra09,
  title = {Quantitative rescattering theory for high-order harmonic generation from molecules},
  author = {Le, Anh-Thu and Lucchese, R. R. and Tonzani, S. and Morishita, T. and Lin, C. D.},
  journal = {Phys. Rev. A},
  volume = {80},
  issue = {1},
  pages = {013401},
  numpages = {23},
  year = {2009},
  month = {Jul},
  publisher = {American Physical Society},
  doi = {10.1103/PhysRevA.80.013401},
  url = {http://link.aps.org/doi/10.1103/PhysRevA.80.013401}
}

@article{Lin:jpb10,
  author={C D Lin and Anh-Thu Le and Zhangjin Chen and Toru Morishita and Robert Lucchese},
  title={Strong-field rescattering physics: self-imaging of a molecule by its own electrons},
  journal={J. Phys. B},
  volume={43},
  number={12},
  pages={122001},
  url={http://stacks.iop.org/0953-4075/43/i=12/a=122001},
  year={2010},
}

@article{Ivanov:jmo05,
author = { Misha Yu   Ivanov  and  Michael   Spanner  and  Olga   Smirnova },
title = {Anatomy of strong field ionization},
journal = {J. Mod. Opt.},
volume = {52},
number = {2-3},
pages = {165-184},
year  = {2005},
publisher = {Taylor & Francis},
doi = {10.1080/0950034042000275360},
}

@article{Spanner:prl03,
  title = {Strong Field Tunnel Ionization by Real-Valued Classical Trajectories},
  author = {Spanner, Michael},
  journal = {Phys. Rev. Lett.},
  volume = {90},
  issue = {23},
  pages = {233005},
  numpages = {4},
  year = {2003},
  month = {Jun},
  publisher = {American Physical Society},
  doi = {10.1103/PhysRevLett.90.233005},
  url = {https://link.aps.org/doi/10.1103/PhysRevLett.90.233005}
}

@article{Lai2015,
  title = {Influence of the Coulomb potential on above-threshold ionization: A quantum-orbit analysis beyond the strong-field approximation},
  author = {Lai, X.-Y. and Poli, C. and Schomerus, H. and Faria, C. Figueira de Morisson},
  journal = {Phys. Rev. A},
  volume = {92},
  issue = {4},
  pages = {043407},
  numpages = {12},
  year = {2015},
  month = {Oct},
  publisher = {American Physical Society},
  doi = {10.1103/PhysRevA.92.043407},
  url = {https://link.aps.org/doi/10.1103/PhysRevA.92.043407}
}

@article{Kay1994_1,
    author = {Kay, Kenneth G.},
    title = "{Integral expressions for the semiclassical time-dependent propagator}",
    journal = {J. Chem. Phys.},
    volume = {100},
    number = {6},
    pages = {4377-4392},
    year = {1994},
    month = {03},
    doi = {10.1063/1.466320},
    url = {https://doi.org/10.1063/1.466320}
}

@article{Walser:jpb03,
doi = {10.1088/0953-4075/36/14/305},
url = {https://dx.doi.org/10.1088/0953-4075/36/14/305},
year = {2003},
month = {jun},
publisher = {},
volume = {36},
number = {14},
pages = {3025},
author = {M W Walser and  T Brabec},
title = {Semiclassical path integral theory of strong-laser-field physics},
journal = {J. Phys. B}
}

@article{Sand:prl99,
  title = {Irregular Orbits Generate Higher Harmonics},
  author = {van de Sand, Gerd and Rost, Jan M.},
  journal = {Phys. Rev. Lett.},
  volume = {83},
  issue = {3},
  pages = {524--527},
  numpages = {0},
  year = {1999},
  month = {Jul},
  publisher = {American Physical Society},
  doi = {10.1103/PhysRevLett.83.524},
  url = {https://link.aps.org/doi/10.1103/PhysRevLett.83.524}
}

@article{Sand:pra00,
  title = {Semiclassical description of multiphoton processes},
  author = {van de Sand, Gerd and Rost, Jan M.},
  journal = {Phys. Rev. A},
  volume = {62},
  issue = {5},
  pages = {053403},
  numpages = {10},
  year = {2000},
  month = {Oct},
  publisher = {American Physical Society},
  doi = {10.1103/PhysRevA.62.053403},
  url = {https://link.aps.org/doi/10.1103/PhysRevA.62.053403}
}

@article{Kay1994_3,
    author = {Kay, Kenneth G.},
    title = "{Semiclassical propagation for multidimensional systems by an initial value method}",
    journal = {J. Chem. Phys.},
    volume = {101},
    number = {3},
    pages = {2250-2260},
    year = {1994},
    month = {08},
    doi = {10.1063/1.467665},
    url = {https://doi.org/10.1063/1.467665}
}

@article{Kay1994_2,
    author = {Kay, Kenneth G.},
    title = "{Numerical study of semiclassical initial value methods for dynamics}",
    journal = {J. Chem. Phys.},
    volume = {100},
    number = {6},
    pages = {4432-4445},
    year = {1994},
    month = {03},
    doi = {10.1063/1.466273},
    url = {https://doi.org/10.1063/1.466273}
}

@article{GreenmanPRA2010,
  title = {Implementation of the time-dependent configuration-interaction singles method for atomic strong-field processes},
  author = {Greenman, Loren and Ho, Phay J. and Pabst, Stefan and Kamarchik, Eugene and Mazziotti, David A. and Santra, Robin},
  journal = {Phys. Rev. A},
  volume = {82},
  issue = {2},
  pages = {023406},
  numpages = {12},
  year = {2010},
  month = {Aug},
  publisher = {American Physical Society},
  doi = {10.1103/PhysRevA.82.023406},
  url = {https://link.aps.org/doi/10.1103/PhysRevA.82.023406}
}

@article{RescignoPRA2000,
  title = {Numerical grid methods for quantum-mechanical scattering problems},
  author = {Rescigno, T. N. and McCurdy, C. W.},
  journal = {Phys. Rev. A},
  volume = {62},
  issue = {3},
  pages = {032706},
  numpages = {8},
  year = {2000},
  month = {Aug},
  publisher = {American Physical Society},
  doi = {10.1103/PhysRevA.62.032706},
  url = {https://link.aps.org/doi/10.1103/PhysRevA.62.032706}
}

@misc{GoetzMolecules,
author = {Goetz, R. E. and Le, A.-T.},
 note = {in preparation}, 
}

@article{BUTCHER1996247,
title = {A history of Runge-Kutta methods},
journal = {Applied Numerical Mathematics},
volume = {20},
number = {3},
pages = {247-260},
year = {1996},
issn = {0168-9274},
doi = {https://doi.org/10.1016/0168-9274(95)00108-5},
url = {https://www.sciencedirect.com/science/article/pii/0168927495001085},
author = {J.C. Butcher}
}

@article{FEAST,
  title = {Density-matrix-based algorithm for solving eigenvalue problems},
  author = {Polizzi, Eric},
  journal = {Phys. Rev. B},
  volume = {79},
  issue = {11},
  pages = {115112},
  numpages = {6},
  year = {2009},
  month = {Mar},
  publisher = {American Physical Society},
  doi = {10.1103/PhysRevB.79.115112},
  url = {https://link.aps.org/doi/10.1103/PhysRevB.79.115112}
}

@book{VMK,
  address = {Singapore},
  author = {Varshalovich, D. A. and Moskalev, A. N. and Khersonskii, V. K.},
  publisher = {World Scientific},
  title = {Quantum Theory of Angular Momentum},
  year = 1988
}

@article{HanPRA2010,
  title = {Comparison between length and velocity gauges in quantum simulations of high-order harmonic generation},
  author = {Han, Yong-Chang and Madsen, Lars Bojer},
  journal = {Phys. Rev. A},
  volume = {81},
  issue = {6},
  pages = {063430},
  numpages = {8},
  year = {2010},
  month = {Jun},
  publisher = {American Physical Society},
  doi = {10.1103/PhysRevA.81.063430},
  url = {https://link.aps.org/doi/10.1103/PhysRevA.81.063430}
}

@article{BandraukPRA2009,
  title = {Quantum simulation of high-order harmonic spectra of the hydrogen atom},
  author = {Bandrauk, A. D. and Chelkowski, S. and Diestler, D. J. and Manz, J. and Yuan, K.-J.},
  journal = {Phys. Rev. A},
  volume = {79},
  issue = {2},
  pages = {023403},
  numpages = {14},
  year = {2009},
  month = {Feb},
  publisher = {American Physical Society},
  doi = {10.1103/PhysRevA.79.023403},
  url = {https://link.aps.org/doi/10.1103/PhysRevA.79.023403}
}

@article{Camp:jpb18,
doi = {10.1088/1361-6455/aaac12},
url = {https://doi.org/10.1088/1361-6455/aaac12},
year = {2018},
month = {feb},
publisher = {IOP Publishing},
volume = {51},
number = {6},
pages = {064001},
author = {Camp, Seth and Beaulieu, Samuel and Schafer, Kenneth J and Gaarde, Mette B},
title = {Resonantly-initiated quantum trajectories and their role in the generation of near-threshold harmonics},
journal = {Journal of Physics B: Atomic, Molecular and Optical Physics},
}

@article{Zagoya:pra12,
  title = {Dominant-interaction Hamiltonians for high-order-harmonic generation in laser-assisted collisions},
  author = {Zagoya, Carlos and Goletz, Christoph-Marian and Grossmann, Frank and Rost, Jan-Michael},
  journal = {Phys. Rev. A},
  volume = {85},
  issue = {4},
  pages = {041401},
  numpages = {4},
  year = {2012},
  month = {Apr},
  publisher = {American Physical Society},
  doi = {10.1103/PhysRevA.85.041401},
  url = {https://link.aps.org/doi/10.1103/PhysRevA.85.041401}
}

@article{Zagoya:njp12,
doi = {10.1088/1367-2630/14/9/093050},
url = {https://doi.org/10.1088/1367-2630/14/9/093050},
year = {2012},
month = {sep},
publisher = {IOP Publishing},
volume = {14},
number = {9},
pages = {093050},
author = {Zagoya, Carlos and Goletz, Christoph-Marian and Grossmann, Frank and Rost, Jan-Michael},
title = {An analytical approach to high harmonic generation},
journal = {New Journal of Physics},
}

@article{Xie:pra23,
  title = {Quantum-corrected semiclassical theory for strong-field ionization},
  author = {Xie, Wenhai and Li, Min and Zhou, Yueming and Lu, Peixiang},
  journal = {Phys. Rev. A},
  volume = {108},
  issue = {6},
  pages = {063102},
  numpages = {10},
  year = {2023},
  month = {Dec},
  publisher = {American Physical Society},
  doi = {10.1103/PhysRevA.108.063102},
  url = {https://link.aps.org/doi/10.1103/PhysRevA.108.063102}
}

@article{Higuet:pra11,
  title = {High-order harmonic spectroscopy of the Cooper minimum in argon: Experimental and theoretical study},
  author = {Higuet, J. and Ruf, H. and Thir\'e, N. and Cireasa, R. and Constant, E. and Cormier, E. and Descamps, D. and M\'evel, E. and Petit, S. and Pons, B. and Mairesse, Y. and Fabre, B.},
  journal = {Phys. Rev. A},
  volume = {83},
  issue = {5},
  pages = {053401},
  numpages = {12},
  year = {2011},
  month = {May},
  publisher = {American Physical Society},
  doi = {10.1103/PhysRevA.83.053401},
  url = {https://link.aps.org/doi/10.1103/PhysRevA.83.053401}
}

@article{Hostetter:pra10,
  title = {Semiclassical approaches to below-threshold harmonics},
  author = {Hostetter, James A. and Tate, Jennifer L. and Schafer, Kenneth J. and Gaarde, Mette B.},
  journal = {Phys. Rev. A},
  volume = {82},
  issue = {2},
  pages = {023401},
  numpages = {8},
  year = {2010},
  month = {Aug},
  publisher = {American Physical Society},
  doi = {10.1103/PhysRevA.82.023401},
  url = {https://link.aps.org/doi/10.1103/PhysRevA.82.023401}
}

@article{Mauger:pra16,
  title = {Semiclassical-wave-function perspective on high-harmonic generation},
  author = {Mauger, Francois and Abanador, Paul M. and Lopata, Kenneth and Schafer, Kenneth J. and Gaarde, Mette B.},
  journal = {Phys. Rev. A},
  volume = {93},
  issue = {4},
  pages = {043815},
  numpages = {11},
  year = {2016},
  month = {Apr},
  publisher = {American Physical Society},
  doi = {10.1103/PhysRevA.93.043815},
  url = {https://link.aps.org/doi/10.1103/PhysRevA.93.043815}
}

@article{Wang:pra21,
  title = {Anomalous ellipticity dependence of the generation of near-threshold harmonics in noble gases},
  author = {Wang, Bincheng and Zhang, Yinfu and Lan, Pengfei and Zhai, Chunyang and Li, Min and Zhu, Xiaosong and Chen, Jing and Lu, Peixiang and Lin, C. D.},
  journal = {Phys. Rev. A},
  volume = {103},
  issue = {5},
  pages = {053119},
  numpages = {11},
  year = {2021},
  month = {May},
  publisher = {American Physical Society},
  doi = {10.1103/PhysRevA.103.053119},
  url = {https://link.aps.org/doi/10.1103/PhysRevA.103.053119}
}

@article{Wang:OptExp24,
author = {Bincheng Wang and Yong Fu and Kan Wang and Zhong Guan and Cheng Jin},
journal = {Opt. Express},
number = {19},
pages = {34034--34049},
publisher = {Optica Publishing Group},
title = {Investigating wavelength dependence of near-threshold harmonics with different atoms and laser intensities: from analysis of an expanded quantum trajectory Monte Carlo model},
volume = {32},
month = {Sep},
year = {2024},
url = {https://opg.optica.org/oe/abstract.cfm?URI=oe-32-19-34034},
doi = {10.1364/OE.533963},
}

@article{MinLi:prl14,
  title = {Classical-Quantum Correspondence for Above-Threshold Ionization},
  author = {Li, Min and Geng, Ji-Wei and Liu, Hong and Deng, Yongkai and Wu, Chengyin and Peng, Liang-You and Gong, Qihuang and Liu, Yunquan},
  journal = {Phys. Rev. Lett.},
  volume = {112},
  issue = {11},
  pages = {113002},
  numpages = {5},
  year = {2014},
  month = {Mar},
  publisher = {American Physical Society},
  doi = {10.1103/PhysRevLett.112.113002},
  url = {https://link.aps.org/doi/10.1103/PhysRevLett.112.113002}
}

@article{Shvetsov:LasPhys25,
doi = {10.1088/1555-6611/ae1e60},
url = {https://doi.org/10.1088/1555-6611/ae1e60},
year = {2025},
month = {dec},
publisher = {IOP Publishing},
volume = {35},
number = {12},
pages = {123001},
author = {Shvetsov-Shilovski, N I},
title = {Trajectory-based models in strong-field physics},
journal = {Laser Physics},
}

@article{Koch:AnnalsPhys21,
title = {A three-step model of high harmonic generation using complex classical trajectories},
journal = {Annals of Physics},
volume = {427},
pages = {168288},
year = {2021},
issn = {0003-4916},
doi = {https://doi.org/10.1016/j.aop.2020.168288},
url = {https://www.sciencedirect.com/science/article/pii/S0003491620302220},
author = {Werner Koch and David J. Tannor},
}

@article{Tannor:AnnuRev00,
   author = "Tannor, David J. and Garashchuk, Sophya",
   title = "Semiclassical Calculation of Chemical Reaction Dynamics via Wavepacket Correlation Functions", 
   journal= "Annual Review of Physical Chemistry",
   year = "2000",
   volume = "51",
   number = "Volume 51, 2000",
   pages = "553-600",
   doi = "https://doi.org/10.1146/annurev.physchem.51.1.553",
   url = "https://www.annualreviews.org/content/journals/10.1146/annurev.physchem.51.1.553",
   publisher = "Annual Reviews",
   issn = "1545-1593",
   type = "Journal Article",
  }

@article{Grossmann:PhysLett98,
title = {From the coherent state path integral to a semiclassical initial value representation of the quantum mechanical propagator},
journal = {Physics Letters A},
volume = {243},
number = {5},
pages = {243-248},
year = {1998},
issn = {0375-9601},
doi = {https://doi.org/10.1016/S0375-9601(98)00265-5},
url = {https://www.sciencedirect.com/science/article/pii/S0375960198002655},
author = {Frank Grossmann and Ademir Luis Xavier},
}

@article{Krause:prl92,
  title = {High-order harmonic generation from atoms and ions in the high intensity regime},
  author = {Krause, Jeffrey L. and Schafer, Kenneth J. and Kulander, Kenneth C.},
  journal = {Phys. Rev. Lett.},
  volume = {68},
  issue = {24},
  pages = {3535--3538},
  numpages = {0},
  year = {1992},
  month = {Jun},
  publisher = {American Physical Society},
  doi = {10.1103/PhysRevLett.68.3535},
  url = {https://link.aps.org/doi/10.1103/PhysRevLett.68.3535}
}

@article{Abanador:jpb17,
doi = {10.1088/1361-6455/50/3/035601},
url = {https://doi.org/10.1088/1361-6455/50/3/035601},
year = {2017},
month = {jan},
publisher = {IOP Publishing},
volume = {50},
number = {3},
pages = {035601},
author = {Abanador, P M and Mauger, F and Lopata, K and Gaarde, M B and Schafer, K J},
title = {Semiclassical modeling of high-order harmonic generation driven by an elliptically polarized laser field: the role of recolliding periodic orbits},
journal = {Journal of Physics B: Atomic, Molecular and Optical Physics},
}

@article{Popruzhenko:pra08,
  title = {Coulomb-corrected quantum trajectories in strong-field ionization},
  author = {Popruzhenko, S. V. and Paulus, G. G. and Bauer, D.},
  journal = {Phys. Rev. A},
  volume = {77},
  issue = {5},
  pages = {053409},
  numpages = {7},
  year = {2008},
  month = {May},
  publisher = {American Physical Society},
  doi = {10.1103/PhysRevA.77.053409},
  url = {https://link.aps.org/doi/10.1103/PhysRevA.77.053409}
}

@article{Zagoya:njp14,
doi = {10.1088/1367-2630/16/10/103040},
url = {https://doi.org/10.1088/1367-2630/16/10/103040},
year = {2014},
month = {oct},
publisher = {IOP Publishing},
volume = {16},
number = {10},
pages = {103040},
author = {Zagoya, C and Wu, J and Ronto, M and Shalashilin, D V and Faria, C Figueira de Morisson},
title = {Quantum and semiclassical phase-space dynamics of a wave packet in strong fields using initial-value representations},
journal = {New Journal of Physics},
}

@article{Burnett:pra92,
  title = {Calculation of the background emitted during high-harmonic generation},
  author = {Burnett, K. and Reed, V. C. and Cooper, J. and Knight, P. L.},
  journal = {Phys. Rev. A},
  volume = {45},
  issue = {5},
  pages = {3347--3349},
  numpages = {0},
  year = {1992},
  month = {Mar},
  publisher = {American Physical Society},
  doi = {10.1103/PhysRevA.45.3347},
  url = {https://link.aps.org/doi/10.1103/PhysRevA.45.3347}
}

@article{Gaarde:jpb08,
  author={Mette B Gaarde and Jennifer L Tate and Kenneth J Schafer},
  title={Macroscopic aspects of attosecond pulse generation},
  journal={Journal of Physics B: Atomic, Molecular and Optical Physics},
  volume={41},
  number={13},
  pages={132001},
  url={http://stacks.iop.org/0953-4075/41/i=13/a=132001},
  year={2008},
}

@article{Popmintchev:science12,
author = {Popmintchev, Tenio and Chen, Ming-Chang and Popmintchev, Dimitar and Arpin, Paul and Brown, Susannah and Alisauskas, Skirmantas and Andriukaitis, Giedrius and Balciunas, Tadas and M\"ucke, Oliver D. and Pugzlys, Audrius and Baltuska, Andrius and Shim, Bonggu and Schrauth, Samuel E. and Gaeta, Alexander and Hernandez-Garcia Carlos and Plaja, Luis and Becker, Andreas and Jaron-Becker, Agnieszka and Murnane, Margaret M. and Kapteyn, Henry C.},
title = {Bright Coherent Ultrahigh Harmonics in the keV X-ray Regime from Mid-Infrared Femtosecond Lasers},
volume = {336},
number = {6086},
pages = {1287-1291},
year = {2012},
doi = {10.1126/science.1218497},
URL = {http://www.sciencemag.org/content/336/6086/1287.abstract},
journal = {Science}
}

@article{Soifer:prl10,
  title = {Near-Threshold High-Order Harmonic Spectroscopy with Aligned Molecules},
  author = {Soifer, H. and Botheron, P. and Shafir, D. and Diner, A. and Raz, O. and Bruner, B. D. and Mairesse, Y. and Pons, B. and Dudovich, N.},
  journal = {Phys. Rev. Lett.},
  volume = {105},
  issue = {14},
  pages = {143904},
  numpages = {4},
  year = {2010},
  month = {Sep},
  publisher = {American Physical Society},
  doi = {10.1103/PhysRevLett.105.143904},
  url = {https://link.aps.org/doi/10.1103/PhysRevLett.105.143904}
}

@article{Botheron:pra09,
  title = {One-electron atom in a strong and short laser pulse: Comparison of classical and quantum descriptions},
  author = {Botheron, P. and Pons, B.},
  journal = {Phys. Rev. A},
  volume = {80},
  issue = {2},
  pages = {023402},
  numpages = {10},
  year = {2009},
  month = {Aug},
  publisher = {American Physical Society},
  doi = {10.1103/PhysRevA.80.023402},
  url = {https://link.aps.org/doi/10.1103/PhysRevA.80.023402}
}

@article{ADK:JETP86,
  author = {Ammosov, M.V. and Delone, N.B. and Krainov, V.P.},
  journal = {JETP},
  volume = {64},
  pages = {1191--1194},
  year = {1986},
}

@article{Le:pra08,
author = {Anh-Thu Le and Toru Morishita and C. D. Lin},
collaboration = {},
title = {Extraction of the species-dependent dipole amplitude and phase from
high-order harmonic spectra in rare-gas atoms},
publisher = {APS},
year = {2008},
journal = {Physical Review A (Atomic, Molecular, and Optical Physics)},
volume = {78},
number = {2},
eid = {023814},
numpages = {6},
pages = {023814},
url = {http://link.aps.org/abstract/PRA/v78/e023814},
doi = {10.1103/PhysRevA.78.023814}
}

@article{Worner:prl09,
author = {Hans Jakob W\"{o}rner and Hiromichi Niikura and Julien B. Bertrand and
P. B. Corkum and D. M. Villeneuve},
collaboration = {},
title = {Observation of Electronic Structure Minima in High-Harmonic Generation},
publisher = {APS},
year = {2009},
journal = {Physical Review Letters},
volume = {102},
number = {10},
eid = {103901},
numpages = {4},
pages = {103901},
url = {http://link.aps.org/abstract/PRL/v102/e103901},
doi = {10.1103/PhysRevLett.102.103901}
}

@article{Tolstikhin:pra11,
  title = {Theory of tunneling ionization of molecules: Weak-field asymptotics including dipole effects},
  author = {Tolstikhin, Oleg I. and Morishita, Toru and Madsen, Lars Bojer},
  journal = {Phys. Rev. A},
  volume = {84},
  issue = {5},
  pages = {053423},
  numpages = {17},
  year = {2011},
  month = {Nov},
  publisher = {American Physical Society},
  doi = {10.1103/PhysRevA.84.053423},
  url = {https://link.aps.org/doi/10.1103/PhysRevA.84.053423}
}

@article{Wahyutama:pra22,
  title = {All-electron, density-functional-based method for angle-resolved tunneling ionization in the adiabatic regime},
  author = {Wahyutama, Imam S. and Jayasinghe, Denawakage D. and Mauger, Fran\ifmmode \mbox{\c{c}}\else \c{c}\fi{}ois and Lopata, Kenneth and Gaarde, Mette B. and Schafer, Kenneth J.},
  journal = {Phys. Rev. A},
  volume = {106},
  issue = {5},
  pages = {052211},
  numpages = {15},
  year = {2022},
  month = {Nov},
  publisher = {American Physical Society},
  doi = {10.1103/PhysRevA.106.052211},
  url = {https://link.aps.org/doi/10.1103/PhysRevA.106.052211}
}

@inproceedings{osg07new,
  title  = {The open science grid},
  author = {
    Pordes, Ruth
    and Petravick, Don
    and Kramer, Bill
    and Olson, Doug
    and Livny, Miron
    and Roy, Alain
    and Avery, Paul
    and Blackburn, Kent
    and Wenaus, Torre
    and W{\"u}rthwein, Frank
    and Foster, Ian
    and Gardner, Rob
    and Wilde, Mike
    and Blatecky, Alan
    and McGee, John
    and Quick, Rob
  },
  doi       = {10.1088/1742-6596/78/1/012057},
  booktitle = {J. Phys. Conf. Ser.},
  volume    = {78},
  series    = {78},
  pages     = {012057},
  year      = {2007},
}

@inproceedings{osg09new,
  title        = {The pilot way to grid resources using glideinWMS},
  author       = {
    Sfiligoi, Igor
    and Bradley, Daniel C
    and Holzman, Burt
    and Mhashilkar, Parag
    and Padhi, Sanjay
    and Wurthwein, Frank
  },
  doi          = {10.1109/CSIE.2009.950},
  booktitle    = {2009 WRI World Congress on Computer Science and Information Engineering},
  volume       = {2},
  series       = {2},
  pages        = {428--432},
  year         = {2009},
}

@misc{osg06new,
  doi = {10.21231/906P-4D78},
  url = {https://osg-htc.org/services/open_science_pool.html},
  author = {{OSG}},
  title = {OSPool},
  publisher = {OSG},
  year = {2006}
}

@misc{osg05new,
  doi = {10.21231/0KVZ-VE57},
  url = {https://osdf.osg-htc.org/},
  author = {{OSG}},
  title = {Open Science Data Federation},
  publisher = {OSG},
  year = {2015}
}

@article{wcl3-x52t,
  title = {Strong-Field Photoelectron Interferometry with Near-Single-Cycle Yb Lasers},
  author = {Hasan, Mahmudul and Tran, Phi-Hung and Gao, Jingsong and Hoang, Van-Hung and Tsai, Ming-Shian and Chen, Ming-Chang and Thumm, Uwe and Cocke, Charles Lewis and Lin, Chii-Dong and Le, Anh-Thu and Han, Meng},
  journal = {Phys. Rev. Lett.},
  volume = {135},
  issue = {26},
  pages = {263001},
  numpages = {7},
  year = {2025},
  month = {Dec},
  publisher = {American Physical Society},
  doi = {10.1103/wcl3-x52t},
  url = {https://link.aps.org/doi/10.1103/wcl3-x52t}
}

@misc{tran2024quantum,
      title={Quantum pathways interference in laser-induced electron diffraction revealed by a semiclassical method}, 
      author={Phi-Hung Tran and Van-Hung Hoang and Anh-Thu Le},
      year={2024},
      eprint={2408.12721},
      archivePrefix={arXiv},
      primaryClass={physics.atom-ph},
      url={https://arxiv.org/abs/2408.12721}, 
}

@misc{mcmanus2025delay,
      title={Delay in electronic vortex states created by multiphoton ionization with single elliptically polarized laser pulses}, 
      author={Edward McManus and Phi-Hung Tran and Michael Davino and Tobias Saule and Van-Hung Hoang and Thomas Weinacht and George Gibson and Anh-Thu Le and Carlos A. Trallero-Herrero},
      year={2025},
      eprint={2509.22817},
      archivePrefix={arXiv},
      primaryClass={physics.atom-ph},
      url={https://arxiv.org/abs/2509.22817}, 
}

@article{Yan:prl2010,
  title = {Low-Energy Structures in Strong Field Ionization Revealed by Quantum Orbits},
  author = {Yan, Tian-Min and Popruzhenko, S. V. and Vrakking, M. J. J. and Bauer, D.},
  journal = {Phys. Rev. Lett.},
  volume = {105},
  issue = {25},
  pages = {253002},
  numpages = {4},
  year = {2010},
  month = {Dec},
  publisher = {American Physical Society},
  doi = {10.1103/PhysRevLett.105.253002},
  url = {https://link.aps.org/doi/10.1103/PhysRevLett.105.253002}
}

@article{Bauer:pra05,
  title = {Strong-field approximation for intense-laser--atom processes: The choice of gauge},
  author = {Bauer, D. and Milo\ifmmode \check{s}\else \v{s}\fi{}evi\ifmmode \acute{c}\else \'{c}\fi{}, D. B. and Becker, W.},
  journal = {Phys. Rev. A},
  volume = {72},
  issue = {2},
  pages = {023415},
  numpages = {5},
  year = {2005},
  month = {Aug},
  publisher = {American Physical Society},
  doi = {10.1103/PhysRevA.72.023415},
  url = {https://link.aps.org/doi/10.1103/PhysRevA.72.023415}
}

@article{Muller:pra99,
  title = {Numerical simulation of high-order above-threshold-ionization enhancement in argon},
  author = {Muller, H. G.},
  journal = {Phys. Rev. A},
  volume = {60},
  issue = {2},
  pages = {1341--1350},
  numpages = {0},
  year = {1999},
  month = {Aug},
  publisher = {American Physical Society},
  doi = {10.1103/PhysRevA.60.1341},
  url = {https://link.aps.org/doi/10.1103/PhysRevA.60.1341}
}

@article{Hehre:jcp69,
    author = {Hehre, W. J. and Stewart, R. F. and Pople, J. A.},
    title = {Self Consistent Molecular Orbital Methods. I. Use of Gaussian Expansions of Slater Type Atomic Orbitals},
    journal = {The Journal of Chemical Physics},
    volume = {51},
    number = {6},
    pages = {2657-2664},
    year = {1969},
    month = {09},
    issn = {0021-9606},
    doi = {10.1063/1.1672392},
    url = {https://doi.org/10.1063/1.1672392}
}

@article{Ditchfield:jcp71,
    author = {Ditchfield, R. and Hehre, W. J. and Pople, J. A.},
    title = {Self Consistent Molecular Orbital Methods. IX. An Extended Gaussian Type Basis for Molecular Orbital Studies of Organic Molecules},
    journal = {The Journal of Chemical Physics},
    volume = {54},
    number = {2},
    pages = {724-728},
    year = {1971},
    month = {01},
    issn = {0021-9606},
    doi = {10.1063/1.1674902},
    url = {https://doi.org/10.1063/1.1674902}
}

@misc{g09,
author="M. J. Frisch and G. W. Trucks and H. B. Schlegel and G. E. Scuseria and M. A. Robb and J. R. Cheeseman and G. Scalmani and V. Barone and B. Mennucci and G. A. Petersson and H. Nakatsuji and M. Caricato and X. Li and H. P. Hratchian and A. F. Izmaylov and J. Bloino and G. Zheng and J. L. Sonnenberg and M. Hada and M. Ehara and K. Toyota and R. Fukuda and J. Hasegawa and M. Ishida and T. Nakajima and Y. Honda and O. Kitao and H. Nakai and T. Vreven and Montgomery, {Jr.}, J. A. and J. E. Peralta and F. Ogliaro and M. Bearpark and J. J. Heyd and E. Brothers and K. N. Kudin and V. N. Staroverov and R. Kobayashi and J. Normand and K. Raghavachari and A. Rendell and J. C. Burant and S. S. Iyengar and J. Tomasi and M. Cossi and N. Rega and J. M. Millam and M. Klene and J. E. Knox and J. B. Cross and V. Bakken and C. Adamo and J. Jaramillo and R. Gomperts and R. E. Stratmann and O. Yazyev and A. J. Austin and R. Cammi and C. Pomelli and J. W. Ochterski and R. L. Martin and K. Morokuma and V. G. Zakrzewski and G. A. Voth and P. Salvador and J. J. Dannenberg and S. Dapprich and A. D. Daniels and  Farkas and J. B. Foresman and J. V. Ortiz and J. Cioslowski and D. J. Fox",
title="Gaussian09 Revision E.01",
note="Gaussian Inc. Wallingford CT 2009"
}

@article{GAMESS,
title = {Recent developments in the general atomic and molecular electronic structure system},
volume = {152},
issn = {0021-9606, 1089-7690},
url = {http://aip.scitation.org/doi/10.1063/5.0005188},
doi = {10.1063/5.0005188},
number = {15},
urldate = {2020-06-18},
journal = {The Journal of Chemical Physics},
author = {Barca, Giuseppe M. J. and Bertoni, Colleen and Carrington, Laura and Datta, Dipayan and De Silva, Nuwan and Deustua, J. Emiliano and Fedorov, Dmitri G. and Gour, Jeffrey R. and Gunina, Anastasia O. and Guidez, Emilie and Harville, Taylor and Irle, Stephan and Ivanic, Joe and Kowalski, Karol and Leang, Sarom S. and Li, Hui and Li, Wei and Lutz, Jesse J. and Magoulas, Ilias and Mato, Joani and Mironov, Vladimir and Nakata, Hiroya and Pham, Buu Q. and Piecuch, Piotr and Poole, David and Pruitt, Spencer R. and Rendell, Alistair P. and Roskop, Luke B. and Ruedenberg, Klaus and Sattasathuchana, Tosaporn and Schmidt, Michael W. and Shen, Jun and Slipchenko, Lyudmila and Sosonkina, Masha and Sundriyal, Vaibhav and Tiwari, Ananta and Galvez Vallejo, Jorge L. and Westheimer, Bryce and Wloch, Marta and Xu, Peng and Zahariev, Federico and Gordon, Mark S.},
year = {2020},
pages = {154102}
}

@article{Tong:pra02,
  title = {Theory of molecular tunneling ionization},
  author = {Tong, X. M. and Zhao, Z. X. and Lin, C. D.},
  journal = {Phys. Rev. A},
  volume = {66},
  issue = {3},
  pages = {033402},
  numpages = {11},
  year = {2002},
  month = {Sep},
  publisher = {American Physical Society},
  doi = {10.1103/PhysRevA.66.033402},
  url = {https://link.aps.org/doi/10.1103/PhysRevA.66.033402}
}
\end{document}